\documentclass[10pt,aps,pre,twocolumn,showpacs,showkeys,groupedaddress,floatfix,longbibliography]{revtex4-1}

\usepackage{graphicx}
\usepackage{amsmath}
\usepackage{amssymb}
\usepackage{hyperref}
\begin{document}

\title[Electrical conductivity of a monolayer]{Electrical conductivity of a monolayer produced by random sequential adsorption of~linear \texorpdfstring{$k$}{k}-mers onto a square lattice}
\author{Yuri Yu. Tarasevich}
\email[Correspondence author: ]{tarasevich@asu.edu.ru}
\affiliation{Astrakhan State University, Astrakhan, Russia}
\author{Valeri  V. Laptev}
\affiliation{Astrakhan State University, Astrakhan, Russia}
\affiliation{Astrakhan State Technical University, Astrakhan, Russia}
\author{Valeria A. Goltseva}
\affiliation{Astrakhan State University, Astrakhan, Russia}
\author{Nikolai I. Lebovka}
\email[Correspondence author: ]{lebovka@gmail.com}
\affiliation{F.D. Ovcharenko Institute of Biocolloidal Chemistry, NAS of Ukraine, Kiev, Ukraine}
\affiliation{Taras Shevchenko Kiev National University, Department of Physics, Kiev, Ukraine}
\date{\today}

\begin{abstract}
The electrical conductivity of a monolayer produced by the random sequential adsorption (RSA) of linear $k$-mers (particles occupying $k$ adjacent adsorption sites) onto a square lattice was studied by means of computer simulation. Overlapping with pre-deposited $k$-mers and detachment from the surface were forbidden. The RSA process continued until the saturation jamming limit, $p_j$. The isotropic (equiprobable orientations of $k$-mers along $x$ and $y$ axes) and anisotropic (all $k$-mers aligned along the $y$ axis) depositions for two different models: of an insulating substrate and conducting $k$-mers (C-model) and of a  conducting substrate and insulating $k$-mers (I-model) were examined. The Frank-Lobb algorithm was applied to calculate the electrical conductivity in both the $x$ and $y$ directions for different lengths ($k=1$ -- $128 $) and concentrations ($p=0$ -- $p_j$) of the $k$-mers.  The `intrinsic electrical conductivity' and concentration dependence of the relative electrical conductivity $\Sigma (p)$ ($\Sigma=\sigma/ \sigma_m$  for the C-model and $\Sigma=\sigma_m /\sigma$ for the I-model, where $\sigma_m$ is the electrical conductivity of substrate) in different directions were analyzed. At large values of $k$ the $\Sigma (p)$ curves became very similar and they almost coincided at $k=128$. Moreover, for both models the greater the length of the $k$-mers the smoother the functions $\Sigma_{xy}(p)$, $\Sigma_{x}(p)$ and $\Sigma_{y}(p)$. For the more practically important C-model, the other interesting findings are
(i)	for large values of $k$ ($k=64, 128$), the values of $\Sigma_{xy}$ and $\Sigma_{y}$ increase rapidly with the initial increase of $p$ from 0 to $0.1$;
(ii) for $k \geq 16$, all the $\Sigma_{xy}(p)$ and $\Sigma_{x}(p)$ curves intersect with each other at the same iso-conductivity points;
(iii) for anisotropic deposition, the percolation concentrations are the same in the $x$ and $y$ directions, whereas, at the percolation point the greater the length of the $k$-mers the larger the anisotropy of the electrical conductivity, i.e., the ratio $\sigma_y/\sigma_x$ ($>1$).
\end{abstract}



\maketitle

\section{Introduction: Electrical conductivity of inhomogeneous media\label{sec:intro}}

The physical properties of inhomogeneous materials (first of all binary materials) have attracted a lot of attention in the scientific community for many decades. This interest is supported by numerous applications such as the production and use of nanocomposites~\cite{McLachlan2007JNM}. Theoretical prediction of the effective properties of such materials  is very important for the  analysis of material performance and for the design of new materials~\cite{Wang2008MSE,Mutiso2015PPS}.

Particular interest is paid to the electrical properties of composites. The theories and models relating to the electrical conductivity, $\sigma$, of mixtures of conducting and insulating species continue to attract great interest from researchers~\cite{Taherian2016CST}.
The limiting case of an infinitely diluted mixture has been extensively considered since the 19th century~\cite{Maxwell1881}.
In a Maxwell approximation~\cite[p.~440--441]{Maxwell1881} and in a Maxwell Garnett approach~\cite{Garnett1904PhylTrans,Garnett1906PhylTrans}, the impurities are supposed to be at a low concentration and have regular compact forms, e.g. sphere or ellipsoid.
For randomly oriented and arbitrarily shaped particles with electrical conductivity $\sigma_p$ suspended in a continuous medium with electrical conductivity $\sigma_m$, the generalized Maxwell model gives the following virial expansion~\cite{Douglas1995AdvChPh,Garboczi1996PRE}
\begin{equation}\label{eq:Garboczi1996}
\frac{\sigma}{\sigma_m} =1+[\sigma]p+ \mathrm{O}(p^2),
\end{equation}
where
\begin{equation}\label{eq:ic}
[\sigma] = \left.\frac{\mathrm{d}\ln\sigma}{\mathrm{d}p}\right|_{p \to 0},
\end{equation}
is called the `intrinsic conductivity', and $p$ is the volume fraction of the particles.
The value of the `intrinsic conductivity' $[\sigma]$ depends upon the electrical conductivity contrast $\Delta=\sigma_p/\sigma_m$, the particle's shape,
the orientation of the particle with respect to the direction of measurement of the electrical conductivity, the spatial dimension $d$,  and the continuous or discrete nature of the problem.
For instance, for randomly distributed hyperspherical particles in $d$ dimensions~\cite{Sangani1990SIAM}
\begin{equation}\label{eq:Sangani1990}
[\sigma]=\frac{d(\Delta-1)}{(\Delta+d-1)}.
\end{equation}
In the two limiting cases, Eq.~\ref{eq:Sangani1990} gives $[\sigma]_\infty=d$ for $\Delta \to \infty$ (conducting inclusions in the insulating medium) and $[\sigma]_0=-d(d-1)$ for $\Delta \to  0^+$ (insulating inclusions in the conducting medium).

By the end of 20th century, the values of $[\sigma]$ for particles with a wide range of shapes had also been estimated~\cite{Douglas1995AdvChPh,Garboczi1996PRE}.
For example, for randomly oriented elliptical inclusions in $d=2$
\begin{equation}\label{eq:Sangani1990sigma}
[\sigma]=\frac{(\Delta^2-1)(1+k)^2}{2(1+k\Delta)(\Delta+k)},
\end{equation}
where $k$ is the ratio of the semi-major to semi-minor axes, i.e., the aspect ratio of the particles. In the two limiting cases, $\Delta\to \infty$  and $\Delta\to 0^+$, this equation gives
\begin{equation}\label{eq:Elliptical1}
[\sigma]_\infty=-[\sigma]_0=1 + (k +k^{-1})/2.
\end{equation}

For the whole composition interval $p \in [0,1]$, a few dozen equations for the concentration dependence of $\sigma(p)$ based on different models had been developed and many comprehensive reviews published~\cite{Clerc1990AdPhys,McLachlan1990JACE,Taherian2016CST}.
Both continuous and discrete models as well as two-dimensional (2D) and three-dimensional (3D) systems have been extensively analyzed to date.

The effective medium approximation (EMA)~\cite{Bruggeman1935AnnPhys} is one of the widely used approaches. The EMA provides a good description of the physical properties at any concentration except the fairly narrow region around the percolation threshold~\cite{Snarskii2007PhysU}.
An alternative description, i.e., the percolation approach, has been applied to a system consisting of randomly distributed conducting and isolating  regions~\cite{Efros1976pssb}. In the percolation approach, the electrical conductivity, $\sigma$, varies with the concentration of the conducting particles, $p$, as $\sigma \propto (p - p_c)^t$, when $p>p_c$, and $\sigma \propto (p_c - p)^{-s}$, when $p<p_c$. Here $p_c$ is the percolation threshold (critical concentration) and, $t$ and $s$ are the critical exponents~\cite{Efros1976pssb}. Note that an extended  approximation obtained in terms of the Maxwell approach allowed description of the electrical properties of the composites for a wide concentration range and even demonstrated the presence of the percolation threshold~\cite{Snarskii2007PhysU}.

Nowadays, the most popular is the so-called generalized effective medium (GEM) equation that accounts for the position of the percolation threshold, $p_c$, and the values of the conductivity exponents $s$, below, and $t$, above, percolation~\cite{McLachlan2007JNM}
\begin{equation}\label{eq:GEM}
(1-p)\frac{\sigma_m^{1/s}-\sigma^{1/s}}{\sigma_m^{1/s}+A\sigma^{1/s}}+p\frac{\sigma_p^{1/t}-\sigma^{1/t}}{\sigma_p^{1/t}+A\sigma^{1/t}}=0,
\end{equation}
where $A=(1-p_c)/p_c$ and $p$ is the concentration of the more conductive species.
At the percolation threshold, $p=p_c$, the GEM equation~\eqref{eq:GEM} gives
\begin{equation}\label{eq:GEMPerc1}
\sigma=\sigma_m^{t/(t+s)}\sigma_p^{s/(t+s)},
\end{equation}
which for a 2D problem reduces to
\begin{equation}\label{eq:GEMPerc2}
\sigma=\sqrt{\sigma_m \sigma_p}.
\end{equation}
This corresponds exactly to the prediction for 2D systems in the case of systems with equal concentrations of the phases $p_c=1/2$~\cite{Dykhne1971JETP}.

For the `intrinsic conductivities', the GEM equation gives
\begin{equation}\label{eq:GEM1}
[\sigma]_\infty=s/p_c
\end{equation}
for the limiting case $\Delta \to \infty$ and
\begin{equation}\label{eq:GEM2}
[\sigma]_0 = -t/(1-p_c)
\end{equation}
for the limiting case $\Delta \to  0^+$.
Note that in the limit of the Bruggeman's symmetric theory (i.e., at $t=s=1$ and $p_c=1/d$~\cite{McLachlan1987JPhC}, the GEM approximation is consistent with the generalized Maxwell model (Eq.~\ref{eq:Sangani1990}) and for 2D systems it gives $[\sigma]_\infty=2$ and $[\sigma]_0= -2$. For 2D random percolation of monomers on a square lattice with $s=t=4/3$ and $p_c=0.5927$~\cite{Stauffer}, the GEM approximation gives
$[\sigma]_\infty\approx 2.25$ and $[\sigma]_0\approx -3.27$.
For 2D systems $s=t=4/3$ and for random percolation of monomers on a square lattice with $p_c=0.5927$~\cite{Stauffer}, the GEM approximation gives
$[\sigma]_\infty\approx 2.25$ and $[\sigma]_0\approx -3.27$.

In many previous experimental and simulation investigations, special interest has been paid to the behaviors of the electrical conductivity and percolation thresholds of the media filled with the particles with anisotropic shapes. Experiments with small conducting carbon rods in an insulating matrix evidenced the strong lowering of the percolation threshold with increased particle length to diameter ratio (aspect ratio)~\cite{Carmona1980JPL}.
The increase in ordering of stick-like carbon black aggregates resulted in an increase in the electrical conductivity anisotropy measured along and perpendicular to the orientation of the aggregates~\cite{Balberg1982APL}. Experiments with graphite platelet-filleds~\cite{Celzard1994SSC} and carbon nanotube-filleds~\cite{Wang2008CST}  nanocomposites revealed the differences in electrical conductivities and percolation thresholds measured along and perpendicular to the orientation of the particles. The effects of nanotube alignment on the percolation conductivity in composites have been studied both experimentally and by Monte Carlo simulations~\cite{Du2005PRB}. The data revealed that the largest conductivity occurred for slightly aligned, rather than isotropic systems.

Note that 2D systems such as metal nanowire films attract particular attention in the scientific community because of their possible applications as flexible, solution-processed transparent conductors~\cite{Mutiso2013PRE,Mutiso2013ACSN}. Computer studies of 2D system of conducting sticks revealed anisotropy  of the electrical conductivity for aligned systems~\cite{Balberg1982APL,Balberg1983SSC}. The general percolation problem of cutting randomly centered insulating holes of arbitrary shape in a 2D conducting sheet and its electrical conductivity has also been investigated~\cite{Garboczi1991PRA}.

Computer simulations have been extensively applied to study percolation and jamming phenomena in oriented and non-oriented 2D systems both for continuous (sticks)~\cite{Balberg1983PRB,Balberg1984PRB,Balberg1987PRA} and for lattice ($k$-mers) problems~\cite{Leroyer1994PRB,Vandewalle2000EPJB,Kondrat2001PRE,Cornette2003epjb,Longone2012PRE,Tarasevich2012PRE,Budinski2016JSM}. For example, with $k$-mers deposited using the random sequential absorption (RSA) model, the data revealed that the percolation threshold has a minimal value when $k \approx 16$ and, probably, percolation is impossible for very long $k$-mers ($k \gtrsim 10^4$)~\cite{Tarasevich2012PRE}. Moreover, defects have a drastic influence on the percolation behavior~\cite{Lebovka2015PRE} of the system of $k$-mers and electrical conductivity of a monolayer produced by aligned $k$-mers~\cite{Tarasevich2015JPhCS}.

However, in spite of the progress in experimental investigations and computer simulations of the electrical properties of 2D composites containing rod-like inclusions~\cite{Mutiso2013PRE,Mutiso2013ACSN}, some issues have not yet been resolved. Of particular interest are the `intrinsic conductivities' and the concentration behavior of the electrical conductivity of systems filled with oriented and non-oriented anisotropic inclusions. Such problems for 2D square lattice systems of $k$-mers deposited using the RSA model have not previously been discussed in the literature.

The rest of the paper is constructed as follows. In Section~\ref{sec:methods}, the technical details of simulations are described, all necessary quantities are defined, and some test results for monomers in comparison with the generalized effective medium approach are presented. Section~\ref{sec:results} presents our principal findings. Section~\ref{sec:conclusion} summarizes the main results.

\section{Methods: Common details of simulation\label{sec:methods}}
In our computer simulation, the random sequential adsorption (RSA) model was used to produce a monolayer~\cite{Evans1993RMP}. The deposition of linear $k$-mers onto a discrete 2D square lattice with periodic boundary conditions (a torus) was performed until a jamming state occurred, i.e., the state when no additional $k$-mers can be placed because the presented voids are too small or of  inappropriate shape. The isotropic as well anisotropic deposition of $k$-mers was examined. During isotropic deposition, both possible orientations of the $k$-mers along the $x$ and $y$ axes are equiprobable. During anisotropic deposition, all the $k$-mers were aligned along the $y$ direction. Overlapping with previously the deposited $k$-mers was strictly forbidden, as a result, a monolayer was formed. Adhesion between deposited $k$-mers and the substrate was assumed to be very strong, so once deposited, a $k$-mer cannot slip over the substrate or leave it (detachment is impossible). We studied the effect of $k$-mer length on the electrical conductivity, $\sigma$,  of the monolayer. The values of $k$ were $2^n$, where  $n=1,2,\dots,7$. Some particular calculations were performed for monomers ($k=1$) in order to make comparisons with the published results.


Two different models were considered:
\begin{itemize}
  \item in the C-model, the electrical conductivity of the $k$-mers $\sigma_p$  is much larger than the electrical conductivity of the bonds of the substrate $\sigma_m$, i.e., $\sigma_p\gg\sigma_m$ ($\Delta \gg 1$, forming conducting inclusions in insulating medium);
  \item in the I-model, the electrical conductivity of the $k$-mers $\sigma_p$  is much smaller than the electrical conductivity of bonds of the substrate $\sigma_m$, i.e., $\sigma_p \ll \sigma_m$ ($\Delta \ll 1$, forming insulating inclusions in conducting medium).
\end{itemize}

\begin{figure}[htbp]
  \centering
  \includegraphics[width=0.75\linewidth]{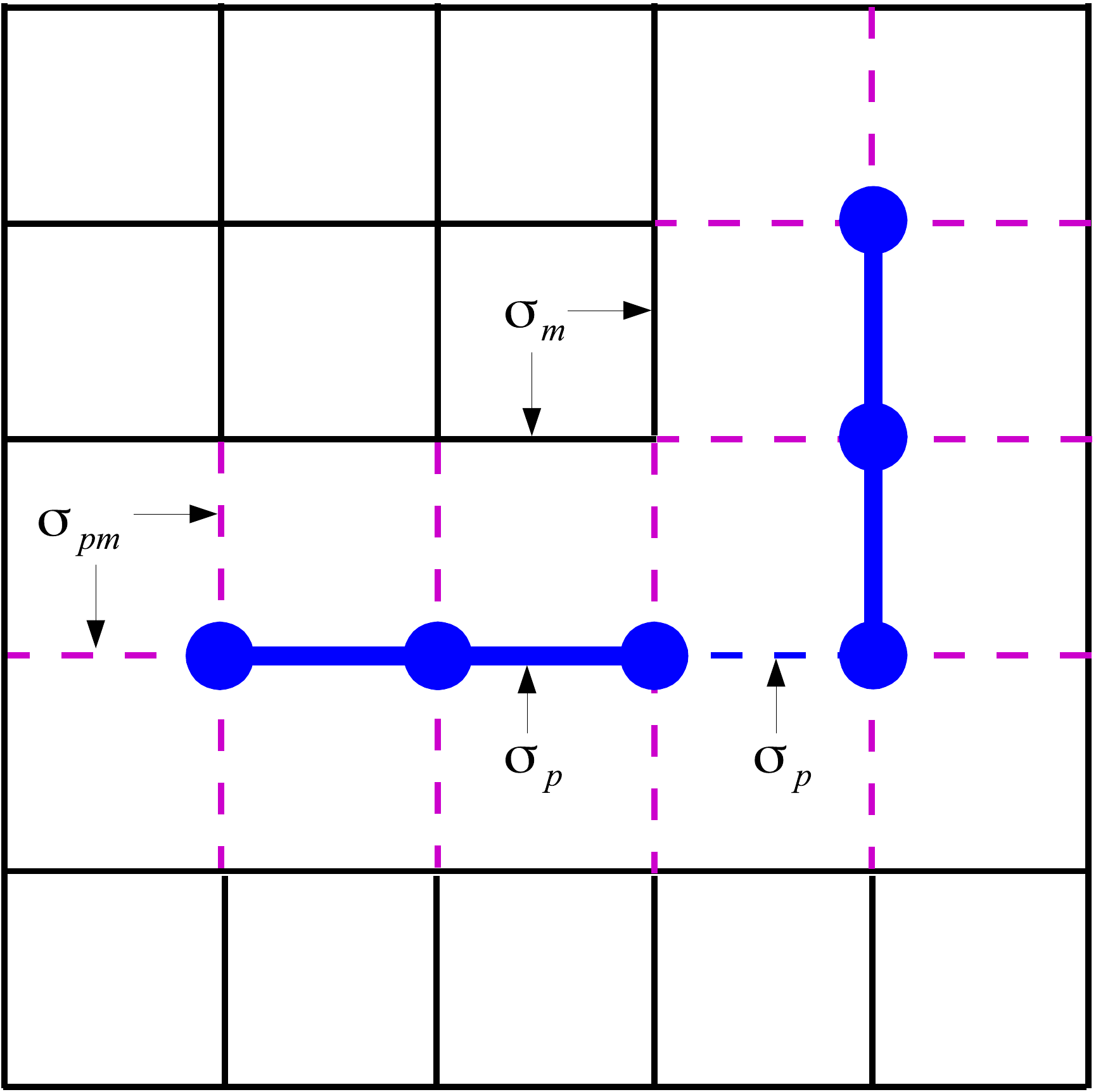}
  \caption{(Color online) Fragment of a square lattice with two deposited 3-mers of different orientations. Conductivities of bonds are indicated.\label{fig:lattice}}
\end{figure}

Different electrical conductivities of the bonds between empty sites, $\sigma_m$, filled sites, $\sigma_p$,  and empty and filled sites, $\sigma_{pm}=2\sigma_p \sigma_m / (\sigma_p+\sigma_m)$ were assumed (Fig.~\ref{fig:lattice}).
For the C-model, we put $\sigma_m =1$, $\sigma_p = 10^6$ in arbitrary units and, for the I-model, we put $\sigma_m =10^6$, $\sigma_p = 1$.

\begin{figure}[htbp]
  \centering
  \includegraphics[width=\linewidth]{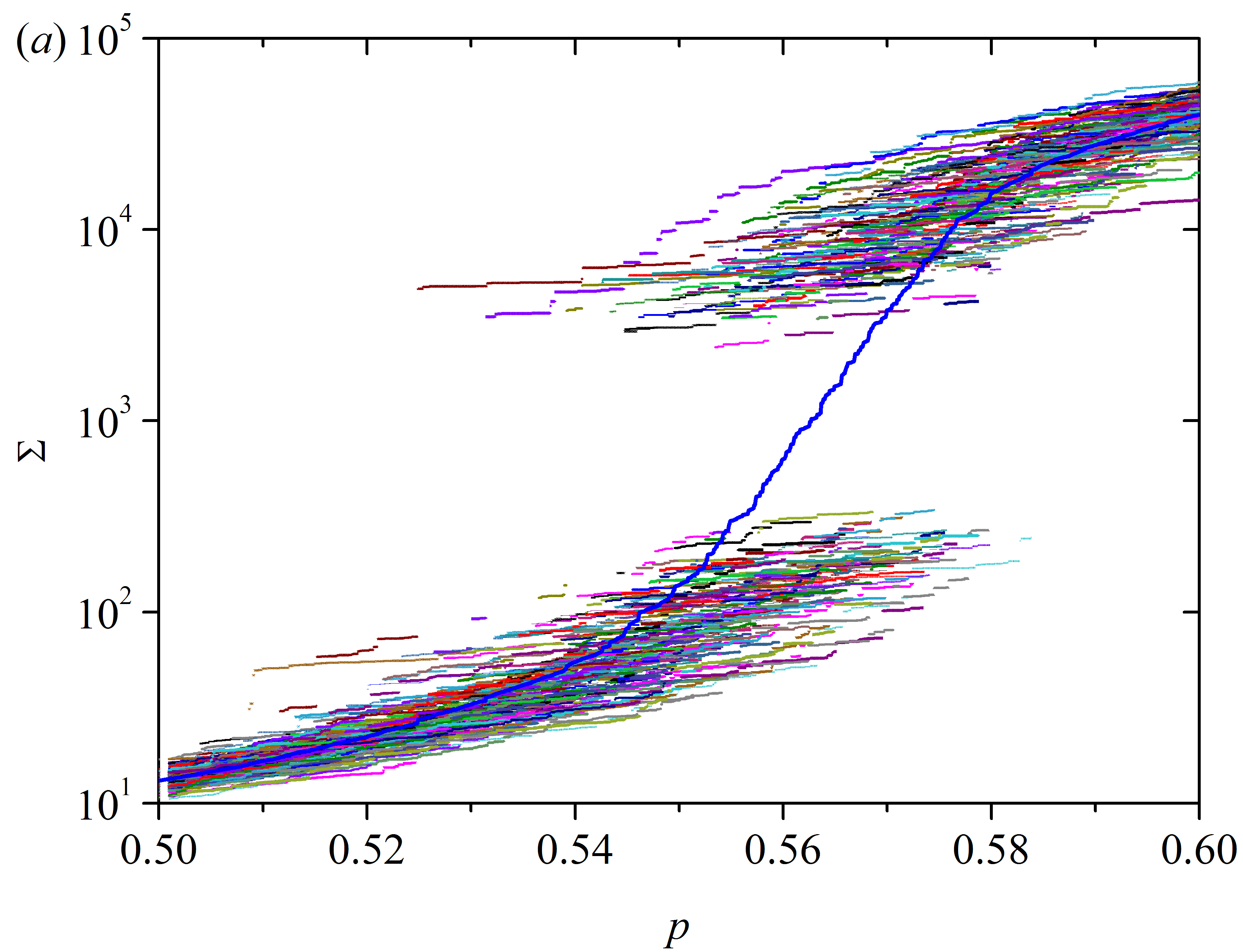}\\
  \includegraphics[width=\linewidth]{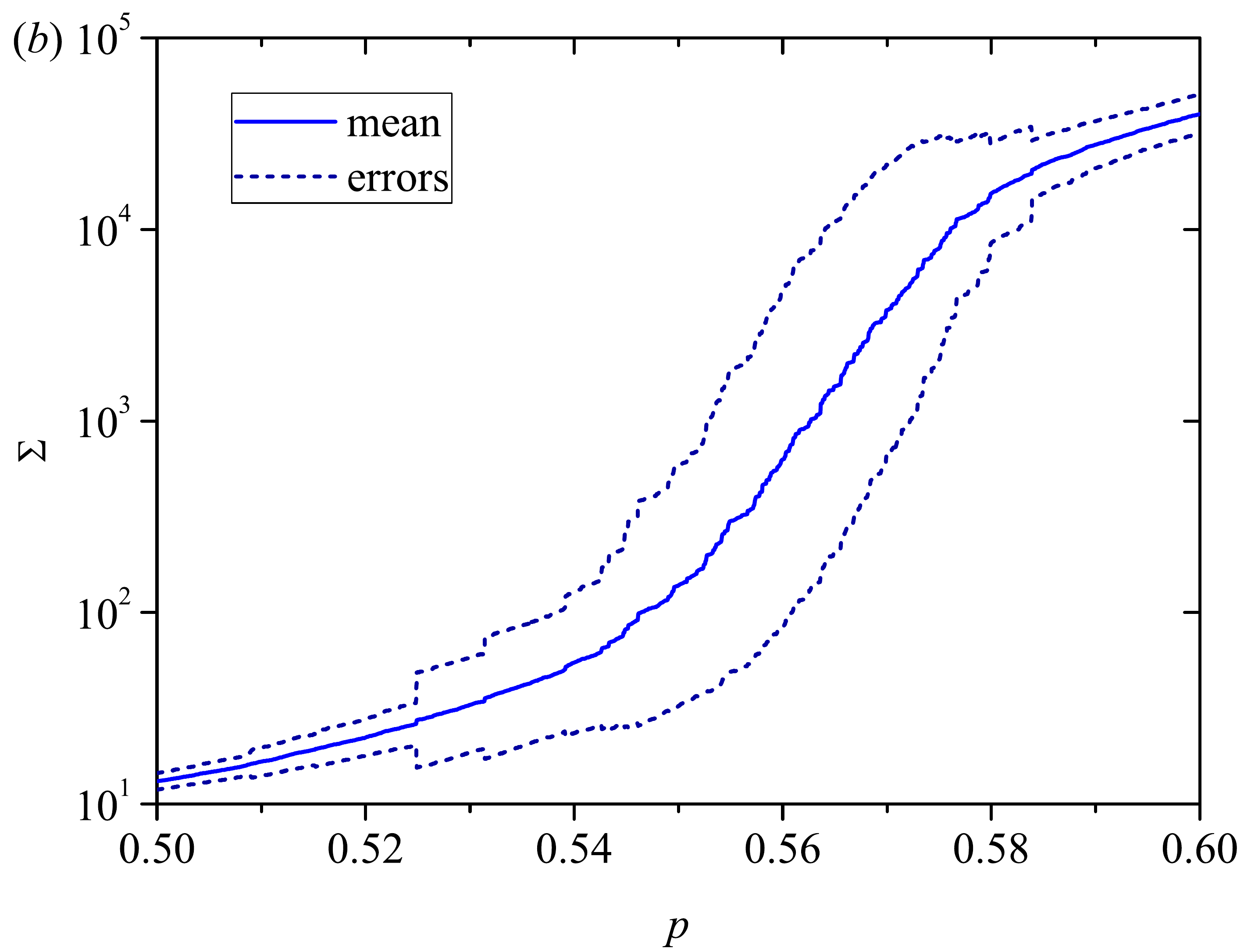}
\caption{(Color online) Relative electrical conductivity $\Sigma$ versus the concentration of $k$-mers, $p$, for $k = 2$, $L = 100k$, isotropic deposition and the C-model.
The enlarged section near the percolation threshold is presented. The diagram shows both the results of 100 runs without averaging (a)~and the corresponding averaged effective conductivity with error bars (b).\label{fig:K2Ave}}
\end{figure}

After deposition of a given number of $k$-mers, the torus was unrolled in a plane and two conducting buses were applied to its  opposite sides. The electrical conductivity was calculated between these buses. The electrical conductivity of the system was calculated using the Frank-Lobb algorithm~\cite{Frank1988PRB} in the $x$ ($\sigma_x$) and $y$ ($\sigma_y$) directions. Note that $\sigma_x$ is the  transversal electrical conductivity, i.e. the electrical conductivity in a  direction perpendicular to the direction of $k$-mer alignment and $\sigma_y$ is the longitudinal electrical conductivity,  i.e. the electrical conductivity along the direction of $k$-mer alignment. The calculations of $\sigma$ were performed each time after the deposition of a given number of particles, until the fraction of occupied lattice sites reached the jamming coverage. For the isotropic case $\sigma_x=\sigma_y=\sigma_{xy}$. The relative conductivity was defined as $\Sigma =\sigma/\sigma_m$ for the C-model and $\Sigma =\sigma_m/\sigma$ for the I-model. For each given value of $k$, the computer experiments were repeated from 10 to 100 times, and then the logarithm of the effective conductivity was averaged. The calculations for the case of $k=128$ and $L=100k$ were too time consuming therefore only one run was performed for this case. Figure~\ref{fig:K2Ave} illustrates the procedure of electrical conductivity averaging for $k = 2$ ($L = 100k$, isotropic deposition and the C-model) using 100 runs. For small values of $k$ ($k<8$), each individual run demonstrates a rather sharp transition from a low conducting state to a high conducting state (Fig.~\ref{fig:K2Ave}a). This jump corresponds to the percolation threshold. Due to the randomness of the deposition of $k$-mers, the value of the percolation threshold may vary a  little between  different runs. The averaging of the logarithms makes the transition smoother (Fig.~\ref{fig:K2Ave}b).
For the example in Fig.~\ref{fig:K2Ave}b (isotropic deposition) the effective conductivity equals $\sqrt{\sigma_m \sigma_p}$ at the percolation transition and this corresponds with the theoretical prediction~\cite{Dykhne1971JETP}.

\begin{figure}[htbp]
  \centering
  \includegraphics[width=\linewidth]{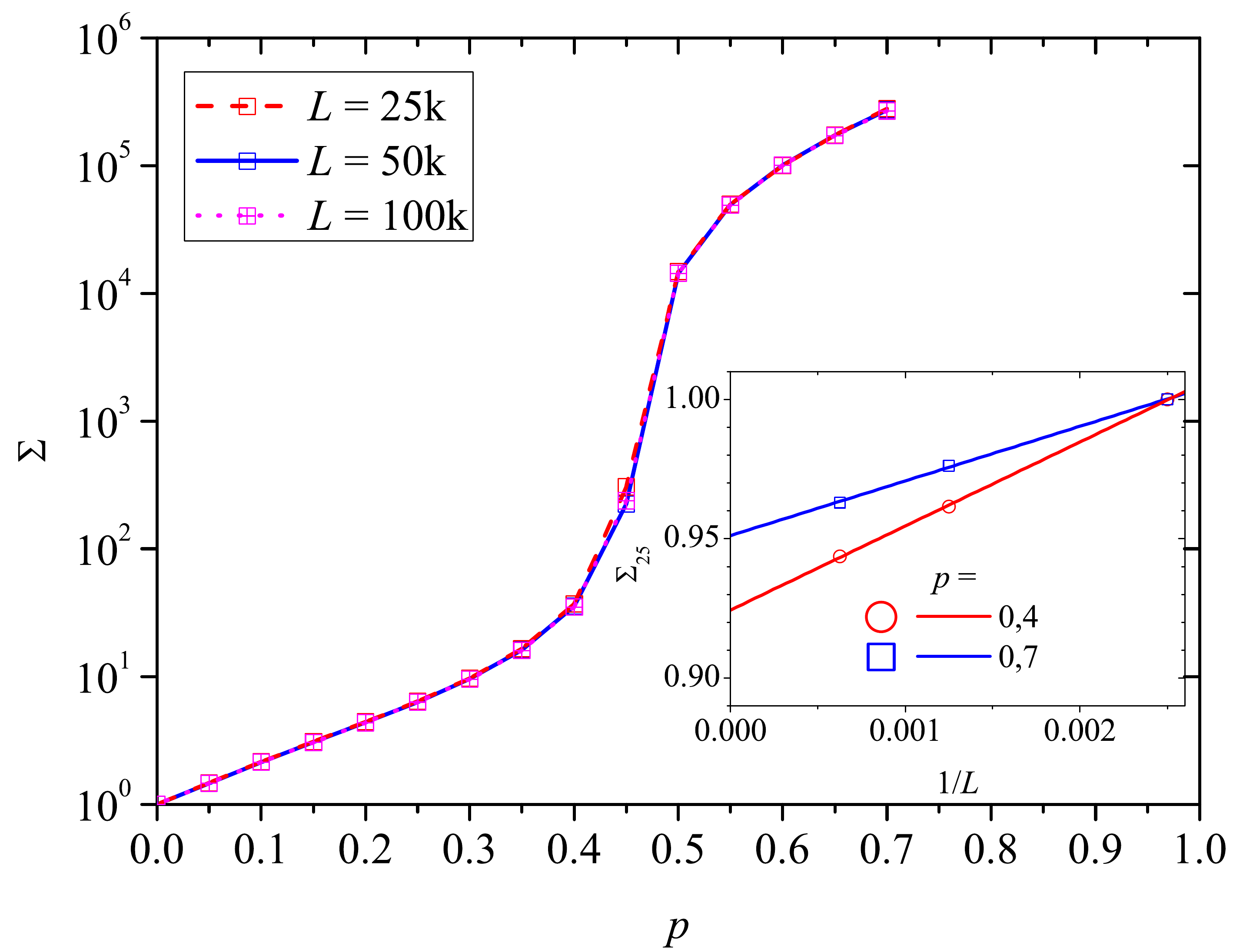}
  \caption{(Color online) Scaling analysis for $k=16$, $L = 25k, 50k, 100k$, the C-model and isotropic deposition.The data are obtained by averaging over 100 independent runs. Inset shows $\Sigma_{25} = \sigma / \sigma_{L=25k}$ versus $1/L$ dependencies at $p = 0.4$ and $p = 0.7$.
  \label{fig:scaling}}
\end{figure}

A scaling analysis of $\sigma(p)$ at different values of $k$ and $L$ was performed. Figure~\ref{fig:scaling} shows an example of the relative electrical conductivity $\Sigma$ versus the concentration of $k$-mers $p$ for the C-model, isotropic deposition, $k=16$, $L=25k, 50k, 100k$.
The difference between the approximated value of electrical conductivity in the limit of the infinite system $\Sigma_{L \to \infty}$ and $\Sigma_{L=100k}$ was of the order of several percents (see, inset to Fig.~\ref{fig:scaling}). This is a reason why in our computations, for any value of $k$, the lattice size $L$ was $L = 100k$.

\begin{figure}[htbp]
  \centering
  \includegraphics[width=\linewidth]{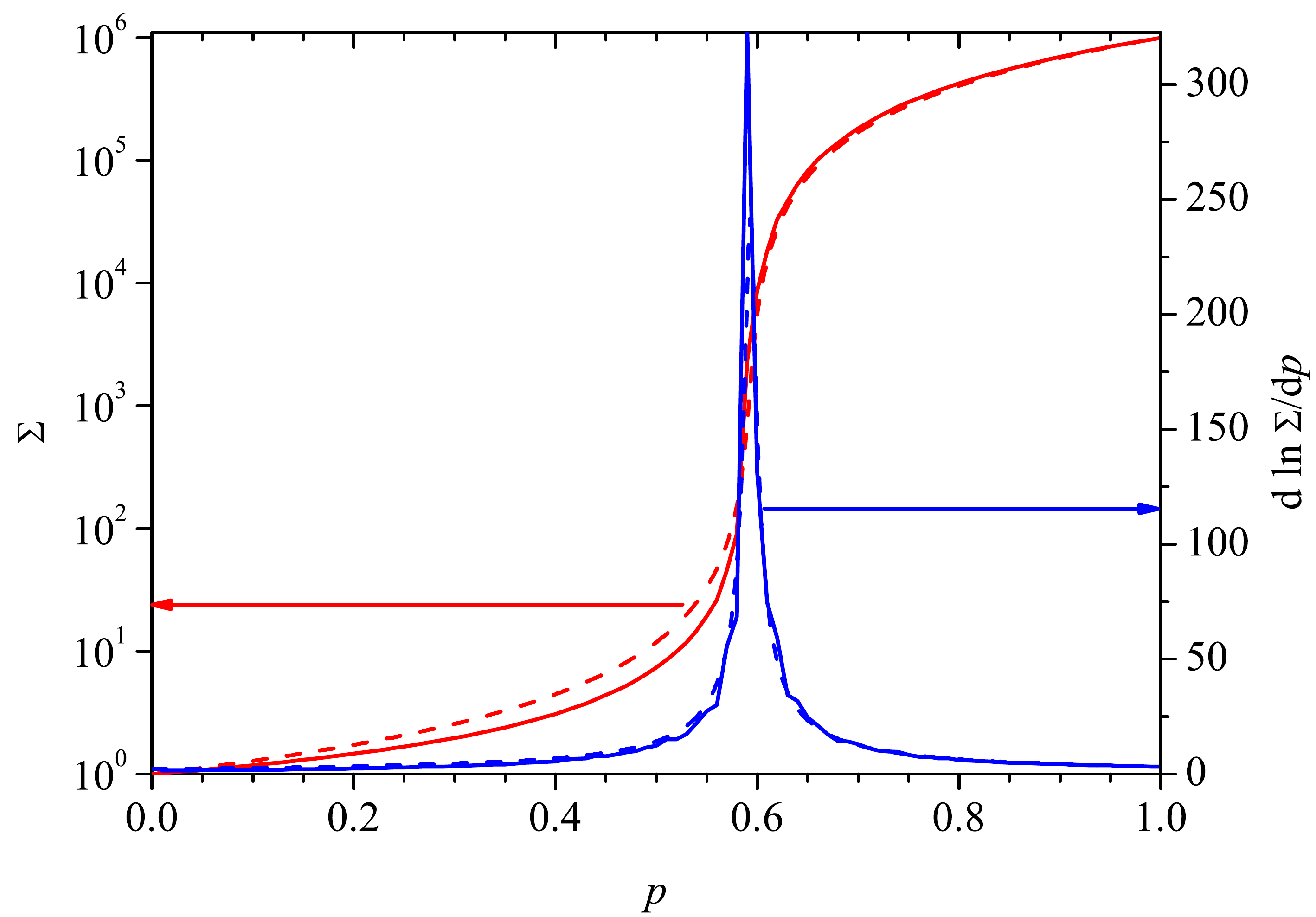}
  \caption{(Color online) Effective conductivity, $\Sigma$, (red) and derivative $\mathrm{d}\ln\Sigma/\mathrm{d}p$ (blue) versus the  concentration of monomers ($k = 1$), $p$, for the C-model, $L=256$. The results are averaged over 1000 runs.  Results obtained from our computer experiments are shown as solid lines. Results of the GEM approximations are shown as dashed lines.\label{fig:Conductivityk1}}
\end{figure}

To estimate the percolation thresholds of the system, the first derivatives of the conductivity plots $\mathrm{d}\ln\Sigma/\mathrm{d}p$ were evaluated. Such an approach is frequently used for determination of the percolation threshold in composite systems~\cite{Celzard1994SSC,Li2007,Li2007CST}. The solid lines in Fig.~\ref{fig:Conductivityk1} give examples of the calculated $\Sigma$ and $\mathrm{d}\ln\Sigma/\mathrm{d}p$ versus the $p$ dependencies for the particular case of monomers ($k=1$). Here, the dashed lines correspond to the prediction of the GEM formula~\eqref{eq:GEM}. In general, the differences between the data from computer simulations and the GEM theory were more noticeable at small concentrations of monomers, $p<p_c$.

\begin{figure}[htbp]
  \centering
  \includegraphics[width=\linewidth]{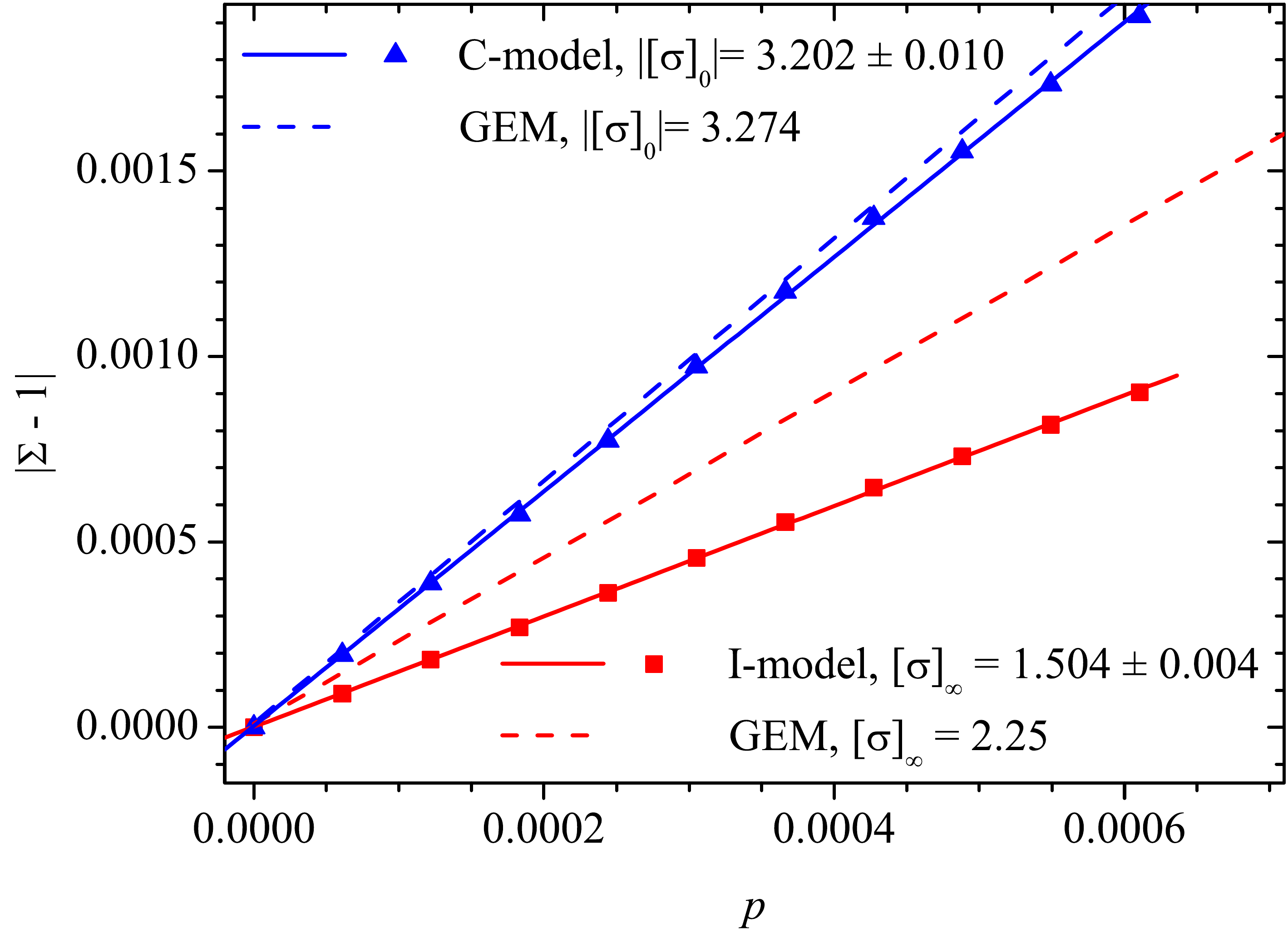}
\caption{(Color online)
Calculated dependencies of the value of $|\Sigma-1|$ versus the concentration of monomers, $p$, for the C-model (red) and the I-model (blue), $L=256$. The results are averaged over 1000 runs. Linear fittings are shown as solid lines. Results of the GEM approximations are shown as dashed lines.\label{fig:ICk1}}
\end{figure}

The `intrinsic conductivity' was calculated from the initial slope of $|\Sigma-1|$ versus the concentration of the $k$-mers $p$ dependencies at small values of $p$. Figure~\ref{fig:ICk1} presents examples of such dependencies for monomers ($k=1$)for the C-model (red) and the I-model (blue). Here, the predictions of the GEM approximation (Eq.~\eqref{eq:GEM}) are also shown by dashed lines. For the C-model the `intrinsic conductivities' were fairly close for the square lattice problem of monomers ($[\sigma]_0=3.202\pm 0.010$) and for the GEM approximation ($[\sigma]_0=3.274$). By contrast, for the I-model, the `intrinsic conductivity' for the square lattice problem of monomers was ($[\sigma]_0=1.504\pm 0.004$) and was noticeably different from that for the GEM approximation ($[\sigma]_0=2.25$).

\section{Results\label{sec:results}}
\subsection{`Intrinsic conductivity'}
\begin{figure}[htbp]
  \centering
  \includegraphics[width=\linewidth]{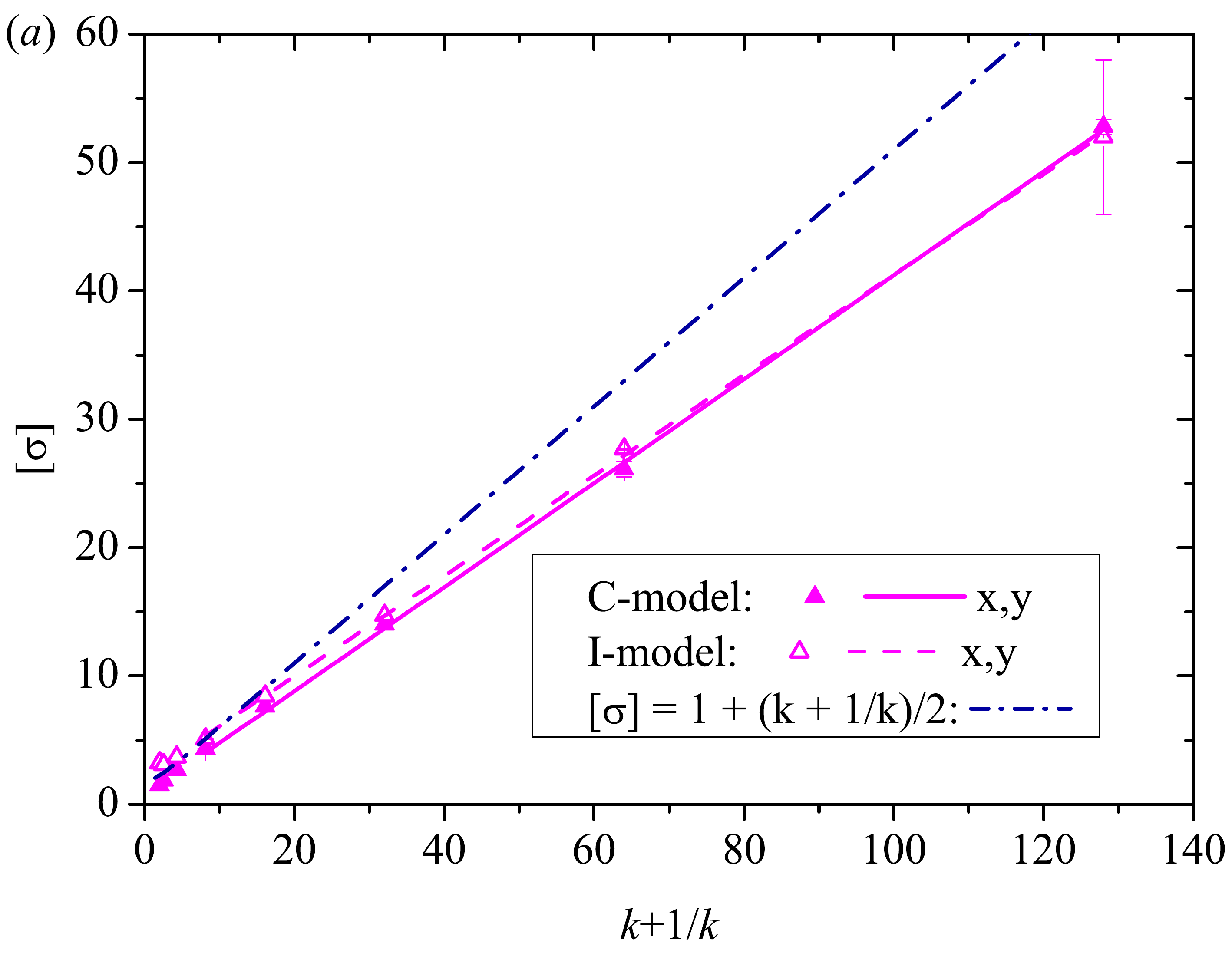}\\
  \includegraphics[width=\linewidth]{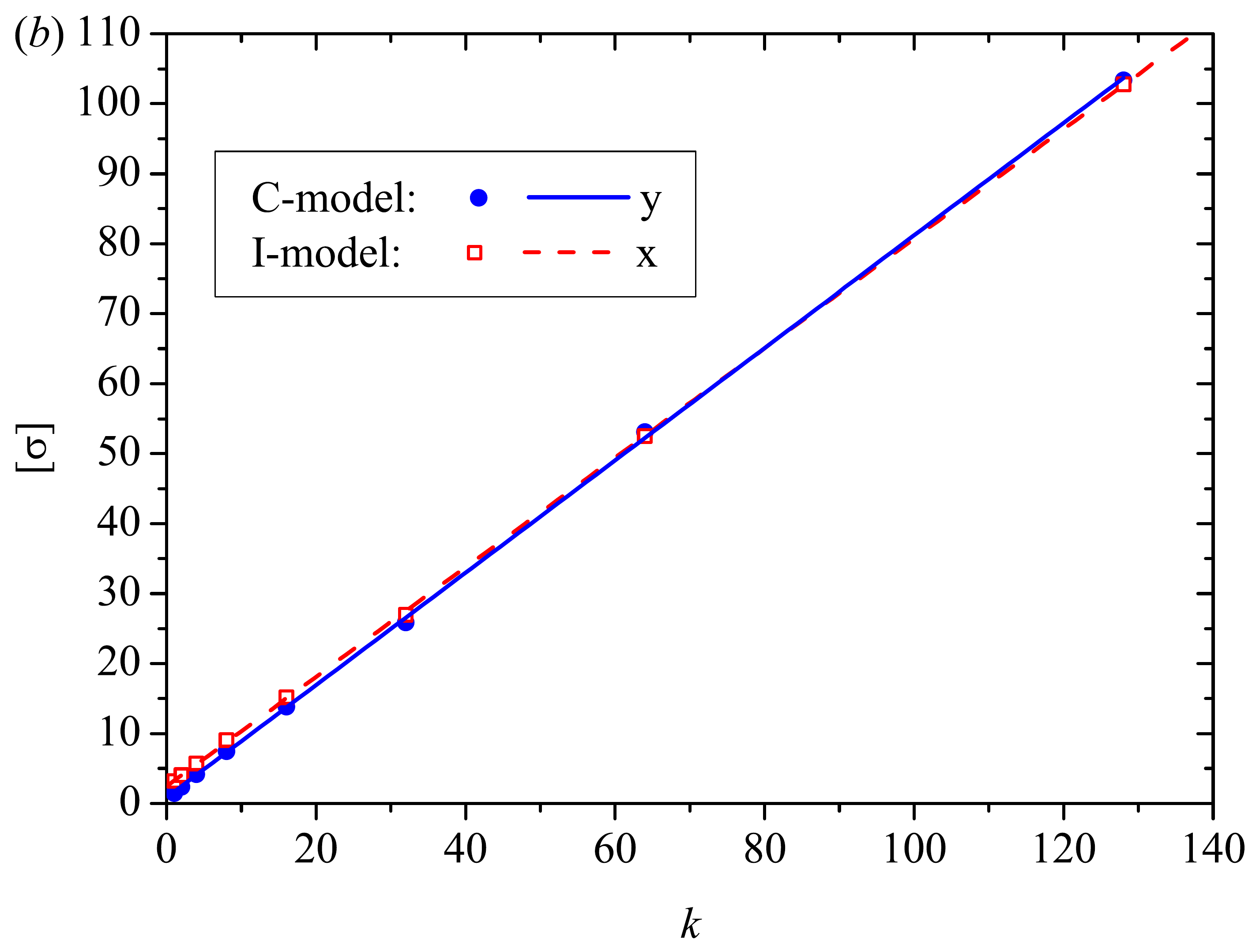}\\
  \includegraphics[width=\linewidth]{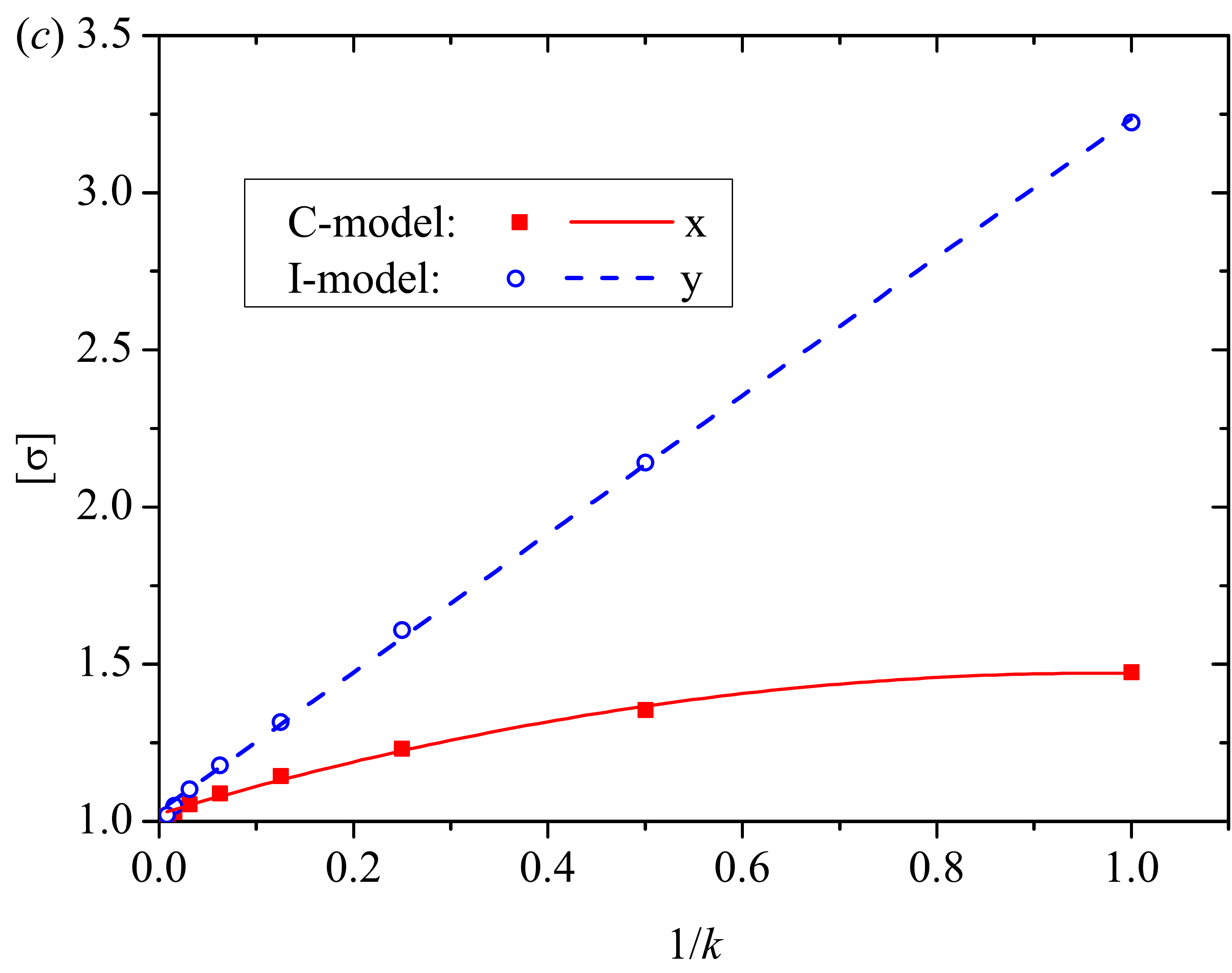}
  \caption{(Color online) `Intrinsic conductivity', $|[\sigma]|$, versus $k$-mer length. $p \to 0$. Calculations and fits. (a)~`Intrinsic conductivity', $|[\sigma_{xy}]|$, versus $k+k^{-1}$, isotropic deposition. The C-model ($R^2=0.998$) and the I-model ($R^2=0.999$). (b)~`Intrinsic conductivity' versus $k$, anisotropic deposition, $[\sigma_x]$ for the C-model,  $|[\sigma_y]|$ for the I-model. (c)~`Intrinsic conductivity' versus $k^{-1}$, anisotropic deposition, $[\sigma_y]$ for the C-model,  $|[\sigma_x]|$ for the I-model.\label{fig:virial1}}
\end{figure}

Figure~\ref{fig:virial1} presents the absolute value of the `intrinsic conductivity' $|[\sigma]|$ versus $k$-mer length for isotropic (a) and anisotropic (b,c) depositions for both the C-model and the I-model.  The values of $\sigma$ are positive for the C-model and negative for the I-model.

For isotropic deposition the value of $|[\sigma]|$ increases as the lengths of the $k$-mers  increases
$$|[\sigma_{xy}]| \propto k + k^{-1}$$.
This linear proportionality was in qualitative correspondence with the prediction of the Maxwell approximation for randomly oriented elliptical inclusions in $d=2$ (Eq.~\eqref{eq:Elliptical1}).

The large differences between the numerical simulations and the predictions of the Maxwell equation may reflect the discreteness of the studied lattice problem for the deposition of $k$-mers.

For anisotropic deposition the behaviors $|[\sigma_x](k)|$ and $|[\sigma_y](k)|$  for the C-model and the I-model were rather different.
For the C-model a linear dependence $[\sigma_y] \propto k$ was observed. However, the $[\sigma_x]$ versus $k^{-1}$ dependence was non-linear. The linear dependencies $\left|[\sigma_x]\right| \propto k$ and $\left|[\sigma_y]\right| \propto k^{-1}$ were only observed for the I-model. The obtained linear dependencies were in qualitative correspondence with the data obtained for perfectly oriented elliptical inclusions in $d=2$~\cite{Garboczi1996PRE}. Table~\ref{tab:intrinsic} summarizes the results for both models.
 \begin{table}[htbp]
 \caption{`Intrinsic conductivities' for the C-model and the I-model at different values of~$k$.\label{tab:intrinsic}}
 \begin{ruledtabular}
 \begin{tabular}{rllllll}
& \multicolumn{3}{c}{C-model} & \multicolumn{3}{c}{I-model} \\
 $k$ & $[\sigma_{xy}]_\infty$ & $[\sigma_x]_\infty$ & $[\sigma_y]_\infty$   & $[\sigma_{xy}]_0$ & $[\sigma_x]_0$ & $[\sigma_y]_0$ \\
  \hline
1   & 2\footnotemark[1]   & --- & ---  & $-2$\footnotemark[1]   & ---  & --- \\
1  & 3.274\footnotemark[2]  & --- & ---  &$-2.25$\footnotemark[3]   & ---  & --- \\ \hline
1   & 1.47  & --- & ---  & $-3.22$  & ---  & --- \\
2   & 1.85  & 1.35 & 2.37  & $-3.09$  & $-4.03$  & $-2.14$\\
4   & 2.68  & 1.23 & 4.15  & $-3.67$  & $-5.71$  & $-1.61$\\
8   & 4.39  & 1.134 & 7.57  & $-5.22$  & $-8.97$  & $-1.33$\\
16  & 7.63  & 1.09 & 13.84 & $-8.44$  & $-15.20$ & $-1.18$\\
32  & 14.02 & 1.05 & 25.90 & $-14.70$ & $-26.94$ & $-1.10$\\
64  & 26.0 & 1.03 & 53.1 & $-28.3$ & $-52.5$ & $-1.05$\\
128 & 52.8 & 1.02 & 103.4 & $-52$ & $-102.8$ & $-1.02$\\
 \end{tabular}
 \end{ruledtabular}
 \footnotetext[1]{Calculated using \eqref{eq:Sangani1990}}
 \footnotetext[2]{Calculated using \eqref{eq:GEM1}}
  \footnotetext[3]{Calculated using \eqref{eq:GEM2}}
 \end{table}

\begin{figure}[htbp]
  \centering
  \includegraphics[width=\linewidth]{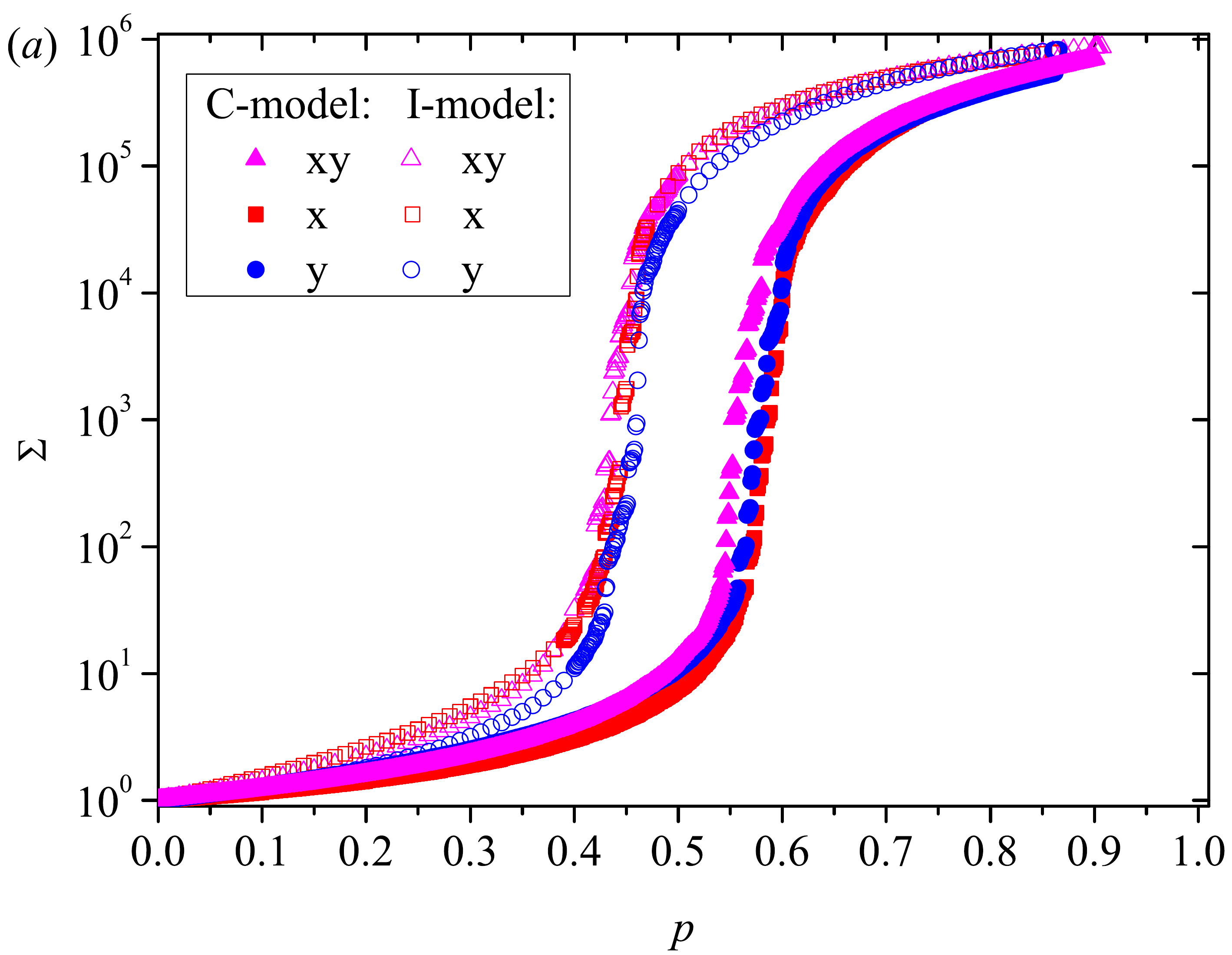}\\
  \includegraphics[width=\linewidth]{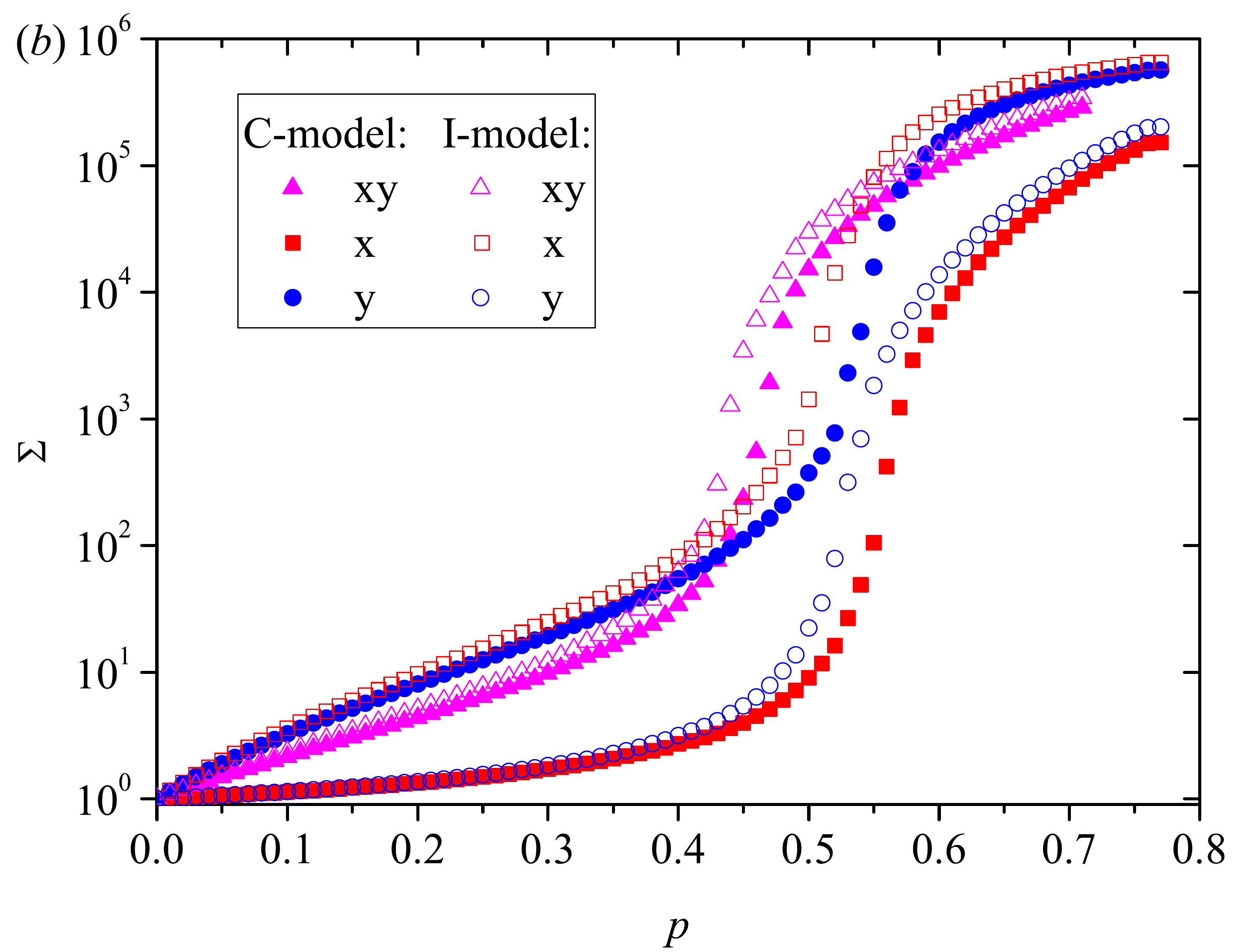}\\
  \includegraphics[width=\linewidth]{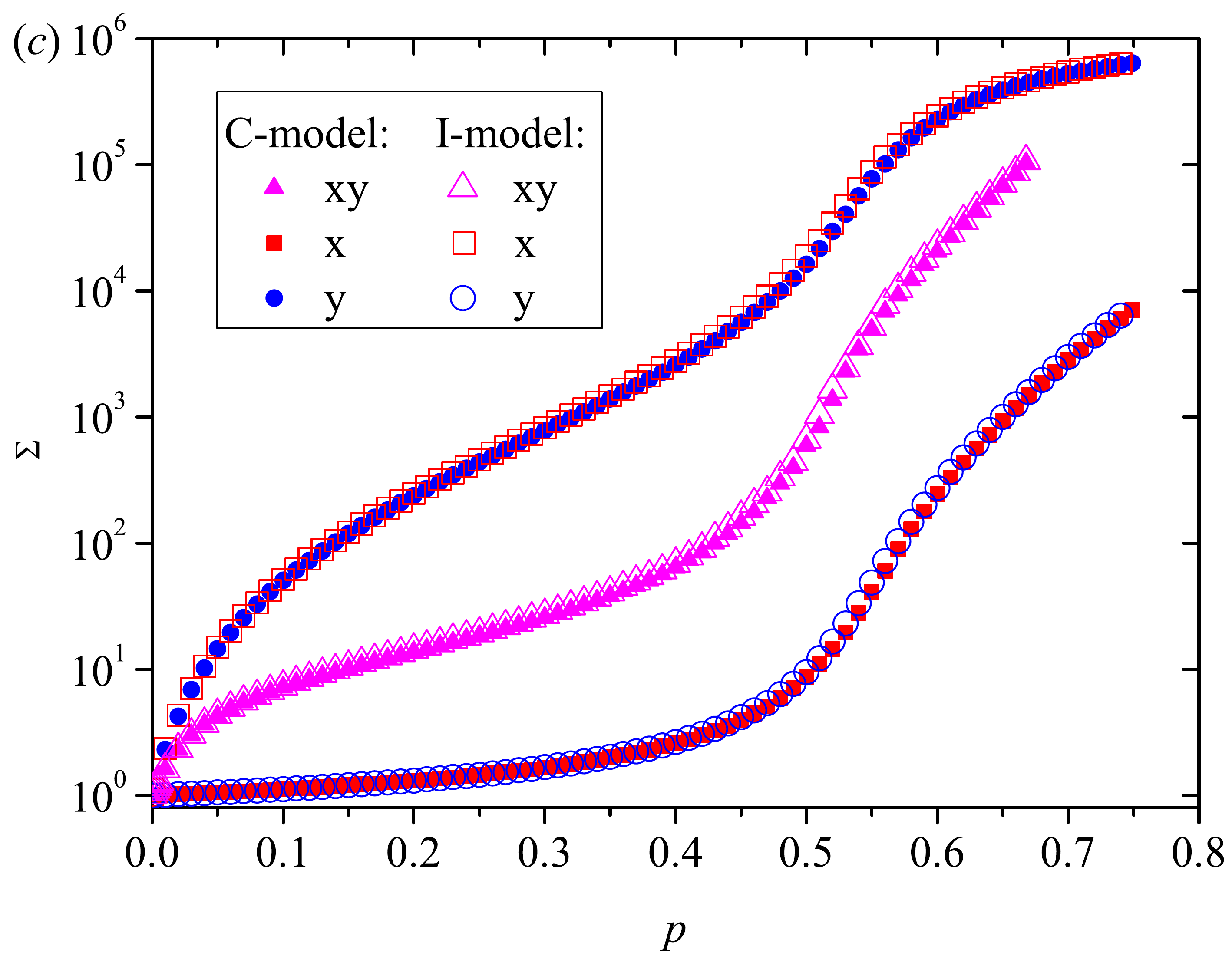}
  \caption{(Color online) Relative electrical conductivity, $\Sigma$, versus concentration of $k$-mers, $p$.  (a)~$k = 2$. (b)~$k = 16$. (c)~$k = 128$. \label{fig:logcondk}}
\end{figure}


\subsection{Effect of \texorpdfstring{$k$}{k}-mer length and anisotropy of deposition on percolation behavior of electrical conductivity}
Figure~\ref{fig:logcondk} compares the dependencies of the relative electrical conductivity, $\Sigma$, versus the concentration of $k$-mers, $p$, for the C-model and the I-model, for (a) $k = 2$, (b) $k = 16$, and (c) $k = 128$. For a small value of $k$ ($k=2$), the percolation transitions are quite sharp, the differences between the electrical conductivity curves for isotropic ($\Sigma_{xy}(p)$) and anisotropic ($\Sigma_{x}(p)$, $\Sigma_{y}(p)$) depositions are fairly small and all the conductivity curves are compactly grouped for both the C-model and the I-model (Fig.~\ref{fig:logcondk}a). For larger values of $k$ ($k \geq 16$), the corresponding differences between $\Sigma_{xy}(p)$, $\Sigma_{x}(p)$, and $\Sigma_{y}(p)$ become significant and the compact grouping for the C-model and the I-model disappears (Fig.~\ref{fig:logcondk}b). Moreover, the greater the length of the $k$-mers the smoother is the percolation transition.
\begin{figure}[htbp]
  \centering
\includegraphics[width=\linewidth]{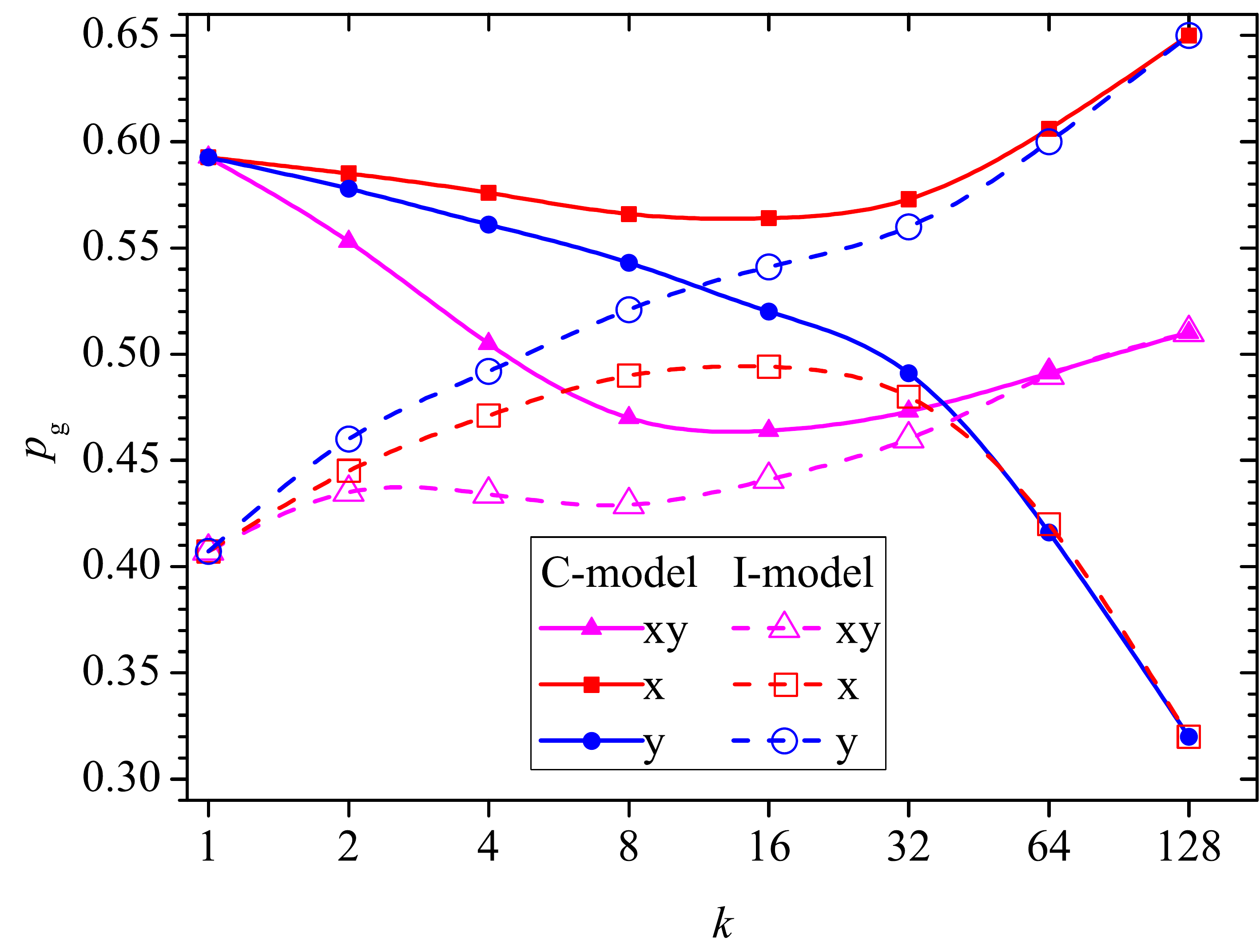}
  \caption{(Color online) Geometrical concentrations $p_g$ versus $k$-mer length ($\log_2$ scale) for the C-model and the I-model evaluated from dependencies $[\sigma_{xy}(p)]$(isotropic deposition) and $[\sigma_{x}(p)]$,$[\sigma_{y}(p)]$ (anisotropic deposition). The lines are provided simply as visual guides.\label{fig:pstar}}
\end{figure}

Finally, at large values of $k$ ($k=128$), another type of compact grouping of the conductivities curves for the C-model and the I-model is observed (Fig.~\ref{fig:logcondk}c). For large values of $k$, the relative electrical conductivities for the C-model and for the I-model agree, within experimental error (Fig.~\ref{fig:logcondk}c). This observation means that at the percolation threshold there is percolation both through empty and through occupied sites.
\begin{figure}[htbp]
  \centering
  \includegraphics[width=\linewidth]{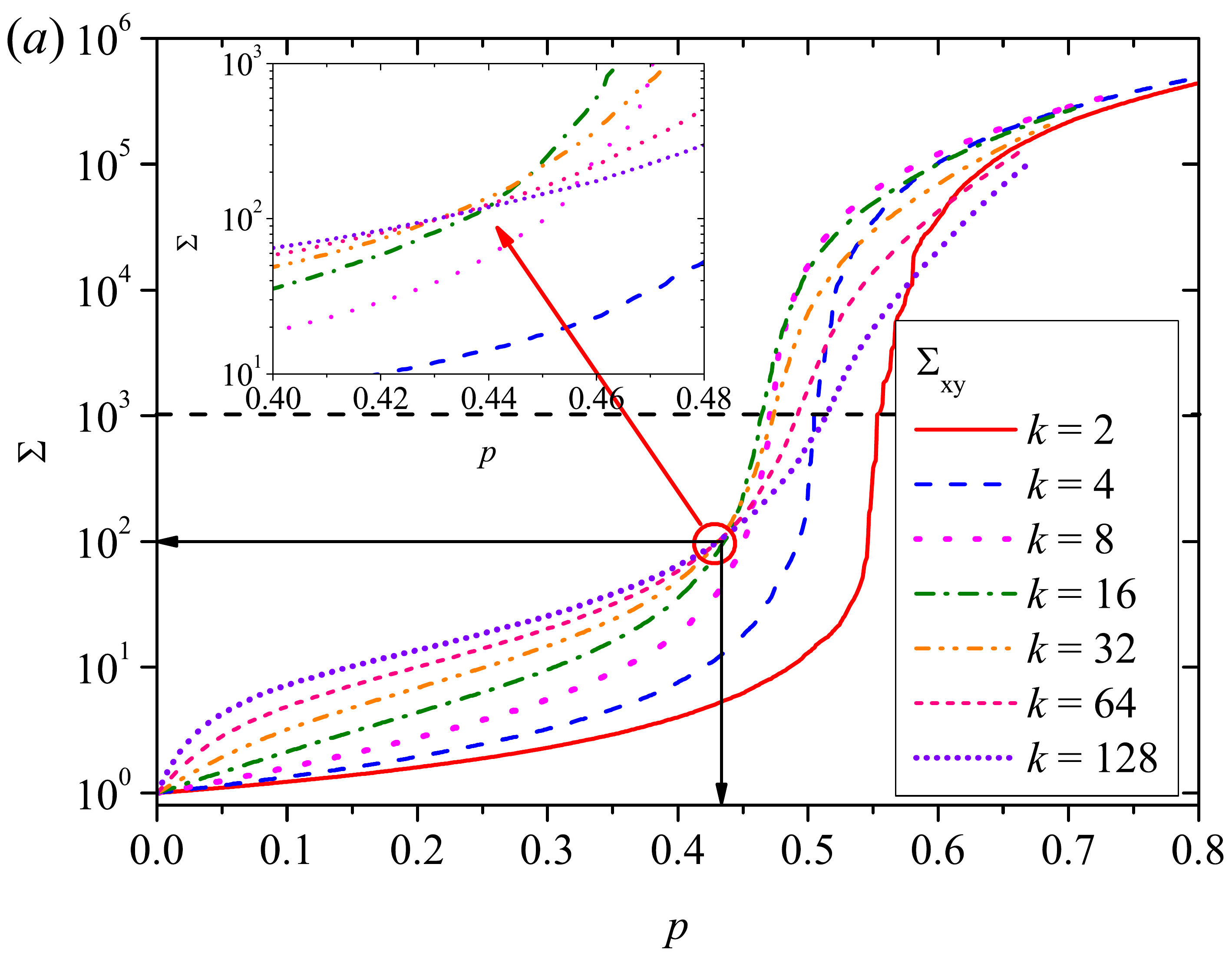}\\
    \includegraphics[width=\linewidth]{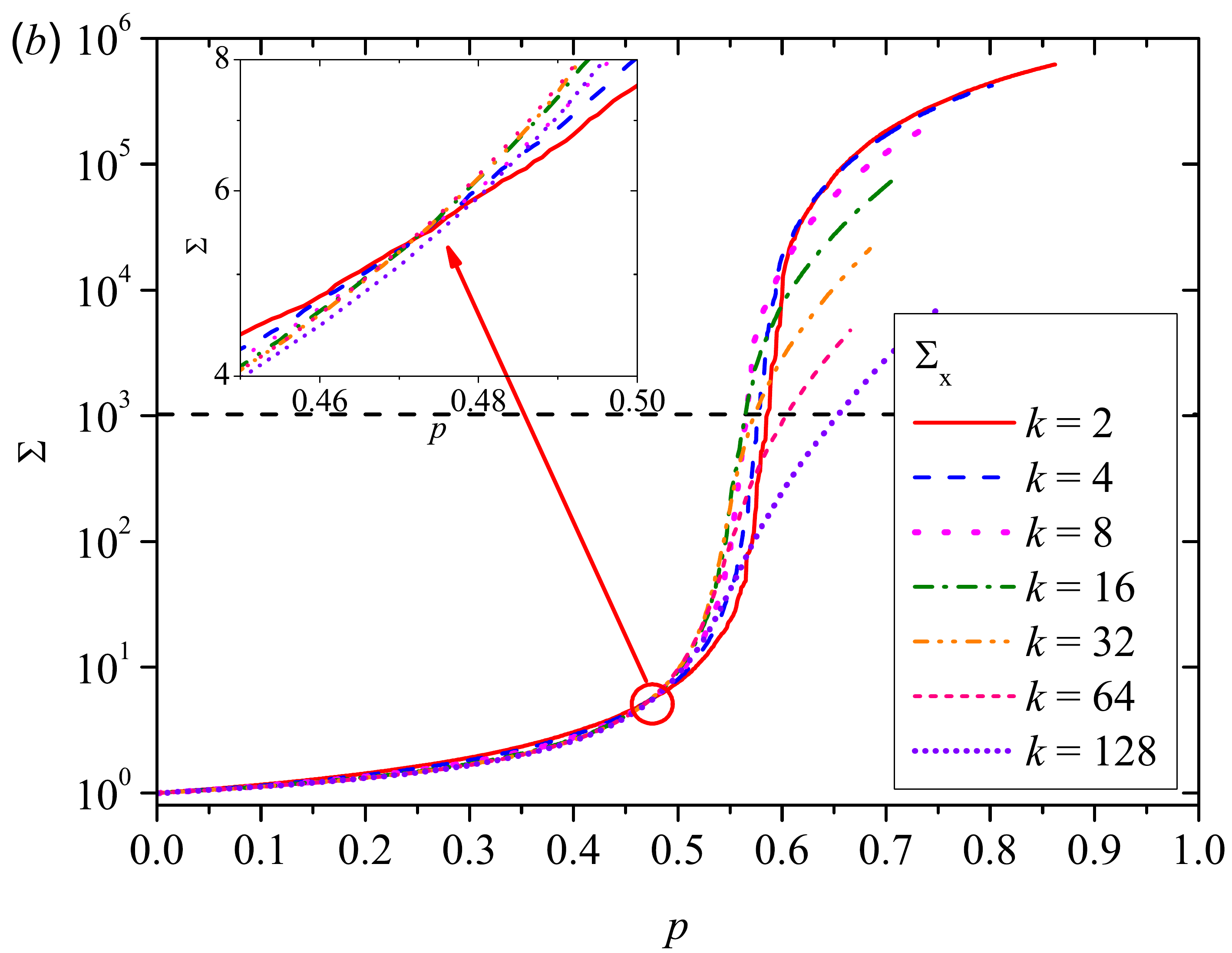}\\
      \includegraphics[width=\linewidth]{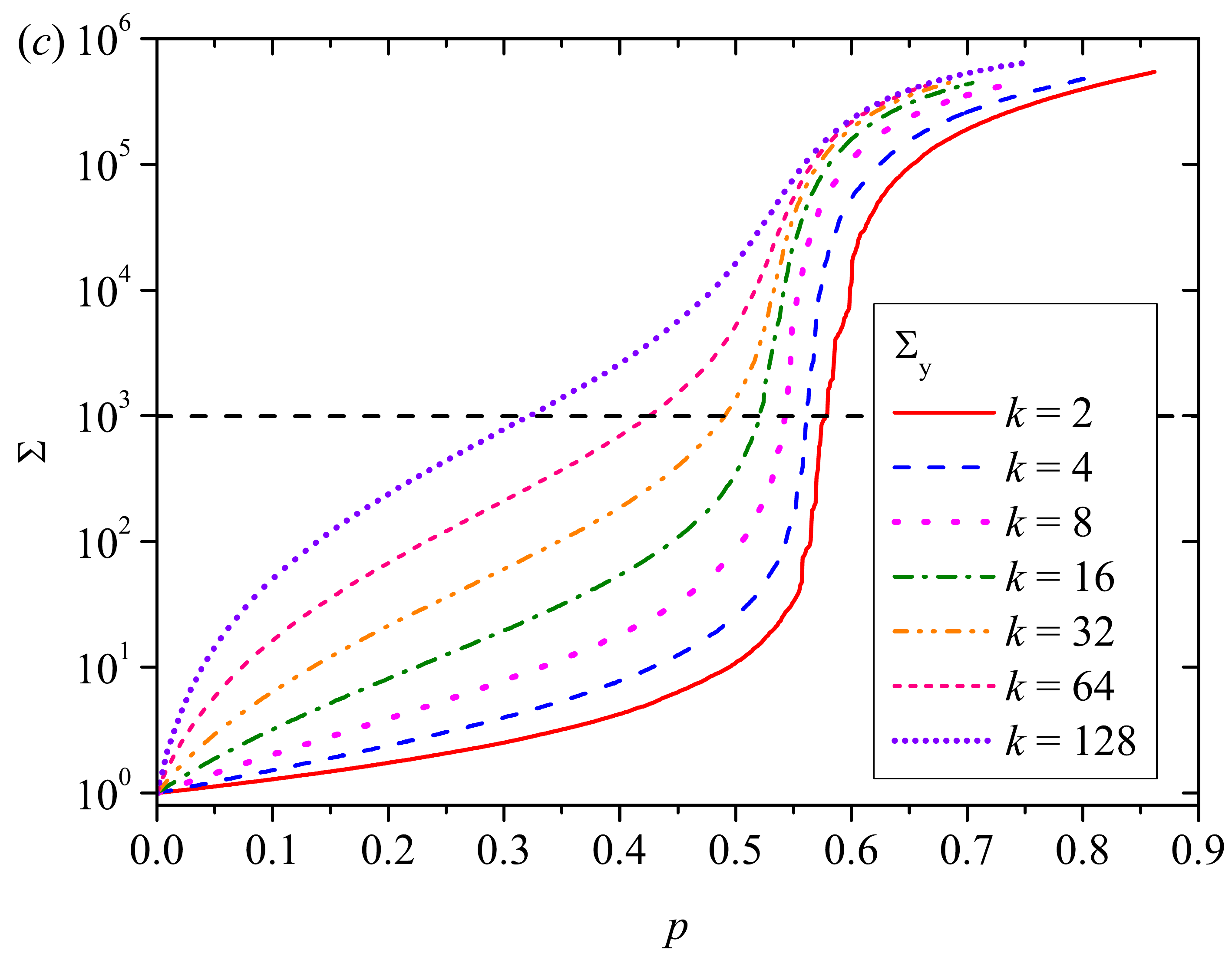}
  \caption{(Color online) Effective conductivity, $\Sigma$ versus concentration of $k$-mers, $p$, for different values of $k$. C-model. (a)~Isotropic deposition. (b)~Anisotropic deposition, transversal conductivity $\Sigma_x$. (c)~Anisotropic deposition, longitudinal conductivity~$\Sigma_y$. \label{fig:allinone}}
\end{figure}

For the sake of clarity, it is useful to analyze the behavior of the geometric concentrations $p_g$ that correspond to the points of mean geometrical conductivity $\Sigma_g=\sqrt{\Sigma_m \Sigma_p}$. Note that the values of $p_g$ are fairly close to the percolation thresholds $p_c$ for isotropic deposition and are in agreement with the theoretical predictions~\cite{Dykhne1971JETP}. For isotropic deposition, the $p_g(k)$ curves go through minima at $k\approx 16$ (C-model) and $k\approx 8$ (I-model) (Figure~\ref{fig:pstar}). This anomaly has previously only been studied in detail for the C-model~\cite{Lebovka2015PRE,Tarasevich2015JPhCS}. The data revealed a minimum of $p_c$ at $k \approx 16$.
For anisotropic deposition, the $p_g(k)$ curves demonstrate in the $x$ direction at $k\approx16$ a minimum  (C-model) or maximum (I-model) whereas the $p_g(k)$ curves in the $y$ direction are monotonic for both models. For large $k$-mers ($k\gtrsim 40$), $p_g^x>p_g^{xy}>p_g^y$ for the C-model and $p_g^y>p_g^{xy}>p_g^x$ for the I-model. The obtained data evidence that anisotropy in percolation behavior depends on the type of model and can be a rather complex function of the length of the $k$-mer. The values of geometric concentrations are presented in Appendix in Table~\ref{tab:pstar}.

\begin{figure}[htbp]
  \centering
  \includegraphics[width=\linewidth]{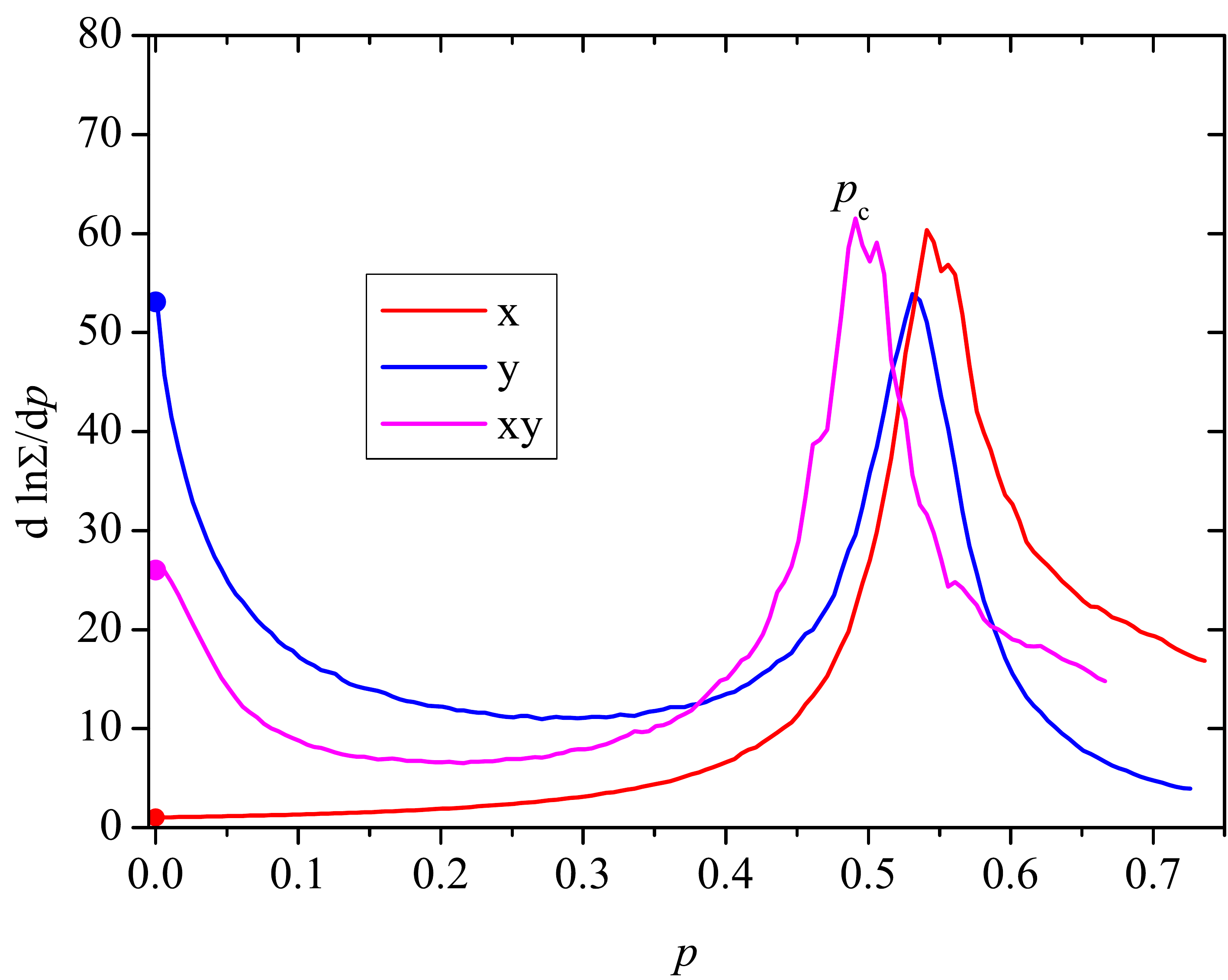}
  \caption{(Color online) Derivative $\mathrm{d}\ln\Sigma/\mathrm{d} p$ versus concentration of $k$-mers, $p$. $k=64$, C-model.
 Circles at $p=0$ correspond to the `intrinsic conductivity', $[\sigma]_0=\mathrm{d}\ln\Sigma/\mathrm{d} p$.
The maxima at $p=p_c$ correspond to the position of the percolation threshold. Compare with Fig.~\ref{fig:Conductivityk1} for monomers.\label{fig:derivatives}}
\end{figure}

In many practically important situations, the electrical conductivity of filler particles is much larger than the electrical conductivity of the host medium, $\sigma_p\gg\sigma_m$ ($\Delta \gg 1$)~\cite{McLachlan2007JNM}. That is why the C-model is the more interesting from a practical point of view and it has been studied in more detail. Figure~\ref{fig:allinone} presents the relative electrical conductivity, $\Sigma$, versus the concentration of $k$-mers, $p$, at different values of $k$ for isotropic deposition (a) and anisotropic deposition (b,c). Figure~\ref{fig:derivatives} presents examples of the calculated logarithmic derivative $\mathrm{d}\ln\Sigma/\mathrm{d}p$ versus the $p$ dependencies for the particular case of $k=64$.

For isotropic deposition, typical $S$-shaped curves $\Sigma_{xy}(p)$ with one
inflection point at the percolation threshold $p=p_c^{xy}$ are observed at small values of $k$ (Fig.~\ref{fig:allinone}a). However, for large values of $k$ ($k=32,64,128$), the $\Sigma_{xy}(p)$ curves demonstrate a second inflection point.  The position of this inflection point corresponds to a minimum at a curve $\mathrm{d}\ln\Sigma/\mathrm{d}p$ (Fig.~\ref{fig:derivatives}).
Clearly expressed maxima at $p=0$ and $p=p_c^{xy}$ are observed at $k \geq 32$.
Visually, these two maxima resemble the two-step percolation transitions typical of composite systems filled by particles with a core-shell structure~\cite{Tomylko2015}.
The magnitude of the maximum at $p=0$ is defined by the `intrinsic conductivity', $\mathrm{d}\ln\Sigma_{xy}/\mathrm{d} p=[\sigma_{xy}]_0$ and it becomes more pronounced at large values of $k$.
A rather sharp transition from the insulating state ($\Sigma =1$) to a relatively conducting state ($\Sigma \approx 10$) is observed when the value of $p$ increases from 0 to $\approx 0.1$. The $S$-shaped curves of $\Sigma_{xy}(p)$ tend to unbend the greater the length of the deposited particles (Fig.~\ref{fig:allinone}).
Surprisingly, for $k\geq 16$, all the curves intersect at one point ($p_i \approx 0.43$, $\Sigma_i \approx 10^2$). This iso-conductivity point may reflect a similarity in the internal structures of the deposits for different values of~$k$.
\begin{figure*}
  \centering
 \includegraphics[width=0.3\linewidth]{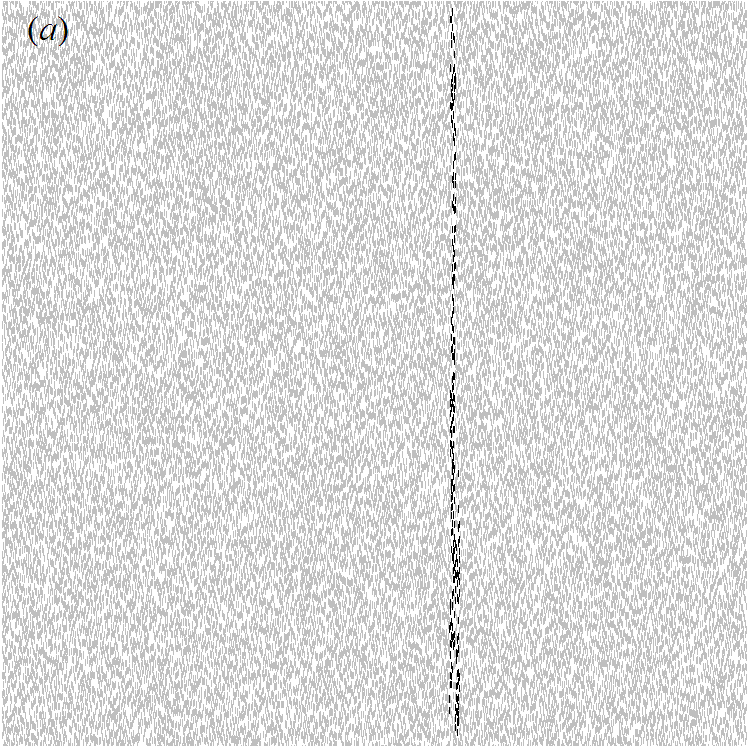}\hfill
 \includegraphics[width=0.3\linewidth]{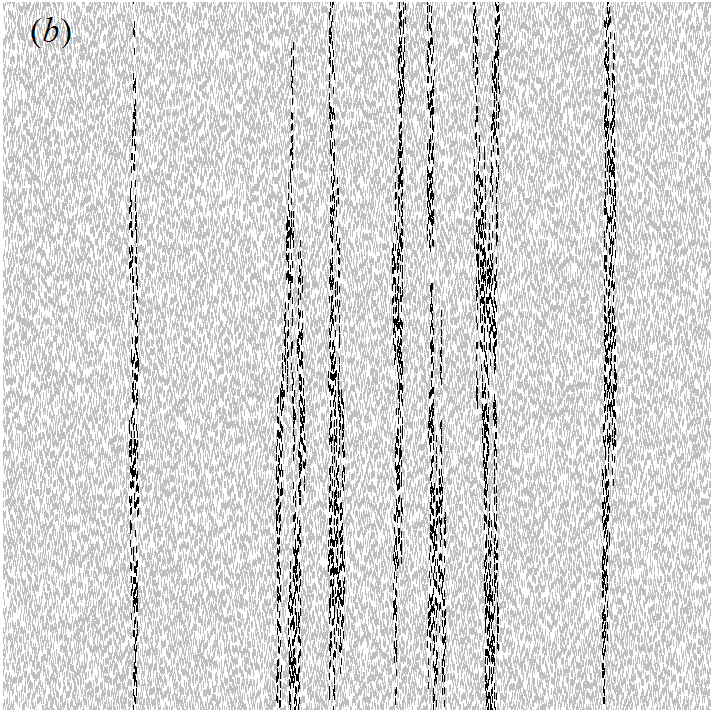}\hfill
 \includegraphics[width=0.3\linewidth]{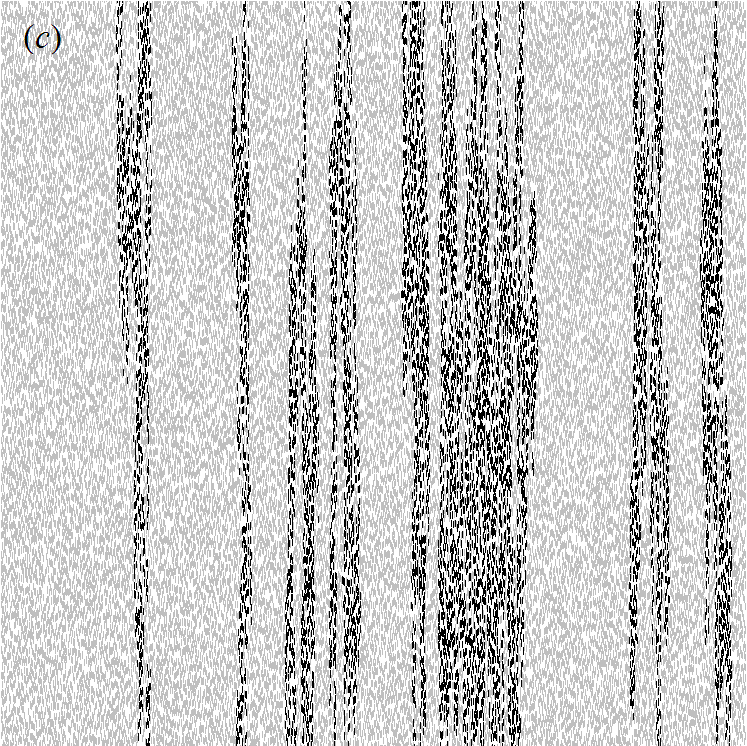}
  \caption{Deposition patterns for different concentrations of $k$-mers, $p$ in the vicinity of the percolation threshold.  $k = 128$, anisotropic deposition, C-model. Empty sites are shown in white, occupied sites are shown in gray, percolation clusters are shown in black. (a)~$p = p_c=0.5118$. One cluster in the vertical direction. (b)~$p = 0.525$. Several clusters in the vertical direction. (c)~$p = 0.53$. More clusters in the vertical direction, some of these clusters are merged in the horizontal direction.\label{fig:cluster}}
\end{figure*}

For anisotropic deposition, an iso-conductivity point ($p_i \approx 0.475$, $\Sigma_i \approx 6$) was observed in the behavior of the transversal conductivity, $\Sigma_{y}(p)$, when $k > 16$ (Fig.~\ref{fig:allinone}b),  whereas the additional inflection point was observed in the behavior of longitudinal conductivity, $\Sigma_{y}(p)$, when $k\geq 8$  (Fig.~\ref{fig:allinone}c). Figure~\ref{fig:cluster}a demonstrates an example of the appearance of the first very narrow and almost linear percolation cluster oriented along the direction of $k$-mer alignment at $p = p_c=0.5118$ ($k=128$, $L=100k$).  Additional percolation clusters of a similar shape arise when the concentration of $k$-mers increases above $p_c$(Fig.~\ref{fig:cluster}b,c). With further increase in the concentration of the deposited $k$-mers, $p$ the almost linear percolation clusters merge and form a spanning structure in both the vertical and horizontal directions. Our previous data of scaling analysis had shown that for infinite systems $p_c^x=p_c^y$ in the studied interval $k=1-128$~\cite{Tarasevich2012PRE}.

For a quantitative description of the anisotropy of the electrical conductivity in the $x$ and $y$ directions, the anisotropy ratio, defined from  $\delta$,\begin{equation}\label{eq:deltaxy}
\sigma_y /\sigma_x =\Delta^\delta,
\end{equation}
was used. $\delta=0$ for isotropic systems and $\delta\approx 1$ for highly anisotropic systems with
$\sigma_y /\sigma_x \approx \Delta$.

\begin{figure}[htbp]
  \centering
\includegraphics[width=\linewidth]{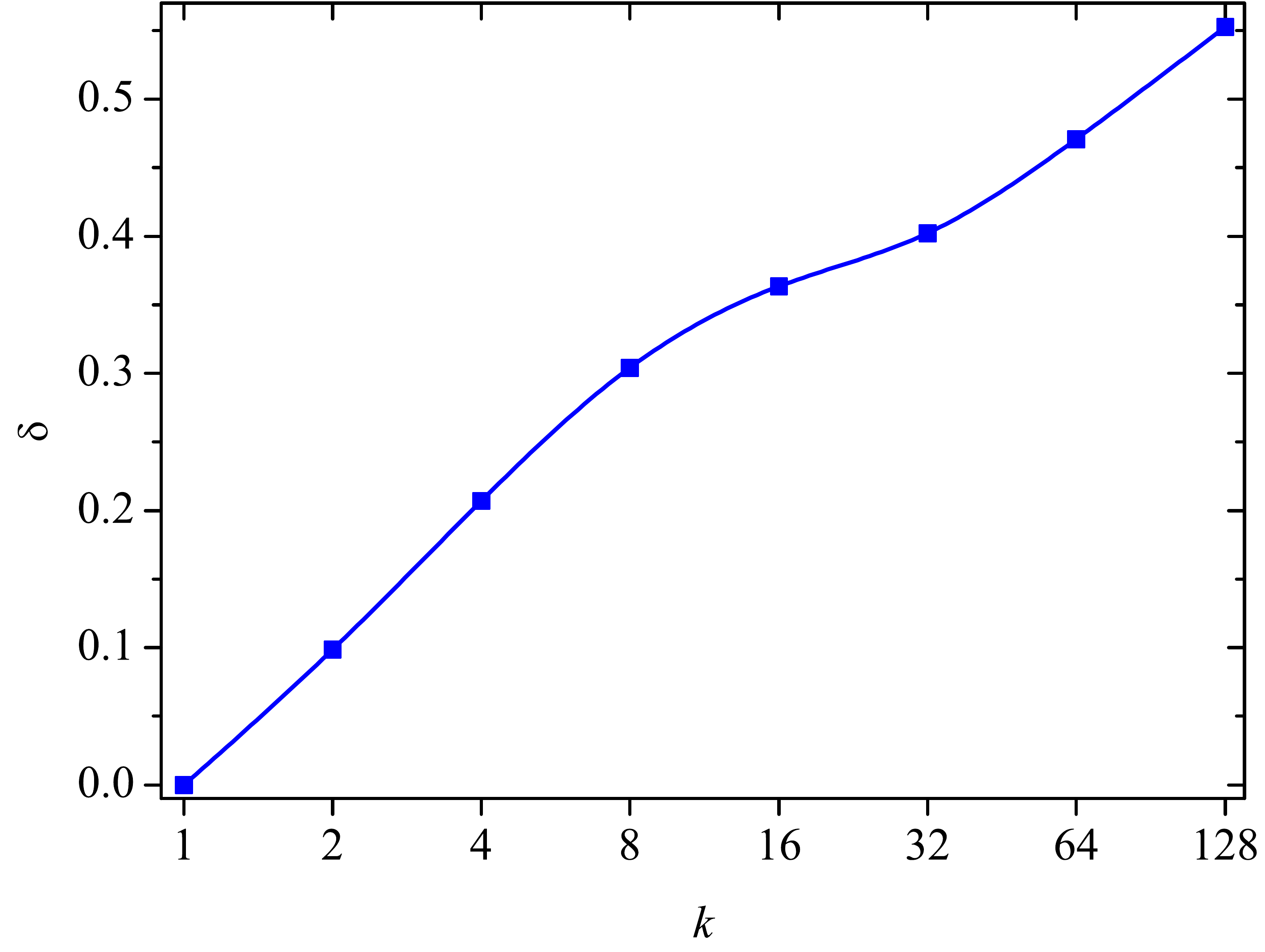}
  \caption{(Color online) Anisotropy of electrical conductivity $\delta$ (Eq.~\eqref{eq:deltaxy}) versus $k$-mer length ($\log_2$ scale). C-model. The line is provided simply as a visual guide.\label{fig:delta}}
\end{figure}

Figure~\ref{fig:delta} presents the anisotropy of electrical conductivity $\delta$ versus the length of the $k$-mers calculated at the threshold concentrations $p_c(k)$ for infinite systems~\cite{Tarasevich2012PRE}. The value of $\delta$ increases with increasing $k$ with some inflection in the $\delta$ versus $\log_2(k)$ dependence at $k\approx 16$ and for very large $k$-mers the ratio of the electrical conductivities $\sigma_y /\sigma_x$ approaches the systems conductivity contrast, $\Delta$. The large difference between $\sigma_y$  and $\sigma_x$ in the percolation point for long $k$-mers definitely reflects the difference in the connectivity along different directions for this anisotropic system.

\section{Conclusion\label{sec:conclusion}}
The behavior of electrical conductivity of a 2D monolayer produced by random sequential adsorption of linear $k$-mers ($k=1$ -- $128$) on a square lattice was analyzed. Two mirrored models were considered: deposition of conducting $k$-mers onto an insulating substrate (C-model) and the embedding of insulating $k$-mers into a conducting substrate (I-model). A large electrical contrast between insulating and conducting species was assumed. Isotropic deposition with two possible orientations of the $k$-mers along the $x$ and $y$ axes and anisotropic deposition with all the $k$-mers oriented in the $y$ direction were examined.

The `intrinsic conductivities' at different values of $k$ were evaluated for the C-model and the I-model.  For isotropic deposition, linear proportionality $|[\sigma_{xy}]| \propto k + k^{-1}$ was observed. For anisotropic deposition, linear dependencies $[\sigma_y] \propto k$ (C-model) and $\left|[\sigma_x]\right| \propto k$, $\left|[\sigma_y]\right| \propto k^{-1}$ (I-model) were observed. The obtained linear dependencies for non-oriented and oriented systems were in qualitative correspondence with the data obtained for elliptical inclusions in $d=2$~\cite{Garboczi1996PRE}. However, $[\sigma_x]$ versus $k^{-1}$ dependence was non-linear for the C-model.

For both models, the sharpness of the electrical conductivity percolation transition decreases with increasing value of $k$. Moreover, at small values of $k$ ($k = 2$), all the relative electrical conductivity curves $\Sigma_{x}(p)$, $\Sigma_{y}(p)$, $\Sigma_{xy}(p)$ are compactly grouped for both the C-model and the I-model.  At large values of $k$ ($k=128$), the compact grouping for both the C- and I-models disappears and the similar relative electrical conductivity curves for the C-model and for the I-models become practically identical.

A more detailed study for the practically interesting the C-model revealed many intriguing behaviors of the relative electrical conductivity, $\Sigma (p)$ for both isotropic and anisotropic depositions. At relatively large values of $k$ ($k=64, 128$), the relative electrical conductivities $\Sigma_{xy}(p)$  and $\Sigma_{y}(p)$ increased rapidly at small concentration of $k$-mers ($p=0$ -- $0.1$) and in the vicinity of the percolation threshold, $p=p_c$. Visually, these resemble two-step percolation transitions. The initial `jump' in relative electrical conductivities reflects the large values of the corresponding `intrinsic conductivities' $[\sigma_{xy}]_0$  and $[\sigma_{y}]_0$.
Surprisingly,  for $k\geq 16$, all the $\Sigma_{xy}(p)$ and $\Sigma_{x}(p)$ curves intersect at the same points ($p_i \approx 0.43$, $\Sigma_i \approx 10^2$ for $\Sigma_{xy}(p)$ curves and $p_i \approx 0.475$, $\Sigma_i \approx 6$ for $\Sigma_{xy}(p)$ curves). At the present time, we have no clear explanation for such behavior. We can speculate that such iso-conductivity points reflect similarities of the internal structure of the deposits at different values of~$k$.
Finally, for large values of $k$ the studied deposits represent unique example of 2D systems with equal percolation thresholds ($p_c^x = p_c^y$) and very different electrical conductivities in the $x$ and $y$ directions ($\Sigma_{y}\gg\Sigma_{x}$).

\section*{Acknowledgements}
The reported study was partially supported by the Ministry of Education and Science of the Russian Federation, Project No.~643 and by the National Academy of Sciences of Ukraine, Project No.~43/16-H (NL).

\appendix
\begin{table}[htbp]
\caption{Geometrical concentrations $p_g$ for the C-model and the I-model at different values of~$k$.\label{tab:pstar}}
\begin{ruledtabular}
\begin{tabular}{rllllll}
& \multicolumn{3}{c}{C-model} & \multicolumn{3}{c}{I-model} \\
$k$ & $p_g^{xy}$ & $p_g^{x}$ & $p_g^{y}$ & $p_g^{xy}$ & $p_g^{x}$ & $p_g^{y}$\\ \hline
1 & 0.5927 & 0.5927 & 0.5927 & 0.4073 & 0.4073 & 0.4073\\
2 & 0.553 & 0.585 & 0.578 & 0.435 & 0.445 & 0.460\\
4 & 0.505 & 0.576 & 0.561 & 0.434 & 0.471 & 0.492\\
8 & 0.470 & 0.566 & 0.543 & 0.429 & 0.490 & 0.521\\
16 & 0.464 & 0.564 & 0.520 & 0.441 & 0.494 & 0.541\\
32 & 0.473 & 0.573 & 0.491 & 0.460 & 0.480 & 0.560\\
64 & 0.491 & 0.606 & 0.416 & 0.490 & 0.420 & 0.600\\
128 & 0.510 & 0.650 & 0.320 & 0.510 & 0.320 & 0.650\\
\end{tabular}
\end{ruledtabular}
\end{table}

\bibliography{conductivity}

\end{document}